\documentclass[lettersize,journal]{IEEEtran}
\usepackage{amsmath,amsfonts}
\usepackage{hyperref}
\usepackage{algorithmic}
\usepackage{algorithm}
\usepackage{array}
\usepackage[caption=false,font=normalsize,labelfont=sf,textfont=sf]{subfig}
\usepackage{textcomp}
\usepackage{stfloats}
\usepackage{url}
\usepackage{verbatim}
\usepackage{graphicx, float}
\graphicspath{{images/}}
\usepackage{cite}
\usepackage{xcolor}
\usepackage[table]{xcolor}
\usepackage{array}
\usepackage{caption}
\usepackage{float}
\usepackage{multirow,graphicx,array} 
\usepackage{multirow,graphicx,array,xcolor,colortbl} 

\begin{document}

\title{\fontsize{24}{28}\selectfont Physics-Guided Conditional Diffusion Networks for Microwave Image Reconstruction}


\author{
    \IEEEauthorblockN{Shirin Chehelgami, Joe LoVetri, and Vahab Khoshdel\IEEEauthorrefmark{1}}\\[1ex]
    \IEEEauthorblockA{
        Department of Electrical and Computer Engineering, University of Manitoba, Winnipeg, Manitoba, Canada\\
        \IEEEauthorrefmark{1}Corresponding author: vahab.khoshdel@umanitoba.ca
    }
}



\pagenumbering{gobble}

\maketitle

\begin{abstract}
A conditional latent-diffusion based framework for solving the electromagnetic inverse scattering problem associated with microwave imaging is introduced. This generative machine-learning model explicitly mirrors the non-uniqueness of the ill-posed inverse problem. Unlike existing inverse solvers utilizing deterministic machine learning techniques that produce a single reconstruction, the proposed latent-diffusion model generates multiple plausible permittivity maps conditioned on measured scattered-field data, thereby generating several potential instances in the range-space of the non-unique inverse mapping. A forward electromagnetic solver is integrated into the reconstruction pipeline as a physics-based evaluation mechanism. The space of candidate reconstructions form a  distribution of possibilities consistent with the conditioning data and the member of this space yielding the lowest scattered-field data discrepancy between the predicted and measured scattered fields is reported as the final solution. Synthetic and experimental labeled datasets are used for training and evaluation of the model. An innovative labeled synthetic dataset is created that exemplifies a varied set of scattering features. Training of the model using this new dataset produces high-quality permittivity reconstructions achieving improved generalization with excellent fidelity to shape recognition. The results highlight the potential of hybrid generative–physics frameworks as a promising direction for robust, data-driven microwave imaging.

\end{abstract}


\section{Introduction}
\IEEEPARstart
{Q}{uantitative} Microwave Imaging (MWI) is a non-invasive technique with applications in a wide range of areas such as medical diagnostics, and non-destructive testing in agricultural and industrial settings  \cite{jimaging8050123, brinker2020review, lovetri2020innovations}. At its core, MWI is a wavefield modality that has associated with it an electromagnetic inverse scattering problem (ISCP). In an electromagnetic ISCP the objective is to quantitatively reconstruct a map of the dielectric properties, and potentially the dielectric loss, of the inaccessible interior of an object or region of interest (OI/ROI). The data used for this reconstruction are the measured scattered fields when the target OI/ROI is interrogated by a known, and usually controllable, impinging incident field. 

The electromagnetic ISCP has associated with it an \textit{ill-posed} wavefield inverse source problem (ISP), that in addition to being highly sensitive to noise in the measurement data, means that more than one permittivity map will produce the same scattered field, \textit{i.e.}, the solution to the ISP is non-unique (see, \textit{e.g.}, \cite{colton1998inverse,devaney2012mathematical}). This makes accurate reconstructions  particularly challenging and usually requires the augmentation of the acquired scattered-field data with some sort of prior information, \textit{e.g.}, the use of regularization methods \cite{colton1998inverse, mojabi2009overview}. A question arises as to whether the non-uniqueness of the ill-posed wavefield inversion problem can be mirrored in an algorithmic procedure to some benefit, as it has, for example, in the formulation of design problems as an inverse problem \cite{Bucci2005, Salucci2022, Brown2020, Palmeri2018, Palmeri2024}.

Traditional algorithms for the ISCP, such as Contrast Source Inversion (CSI)~\cite{vandenberg1997contrast} and Gauss-Newton Inversion \cite{DeZaeytijd2007, rubaek2007nonlinear}, rely on iterative optimization. These approaches typically incorporate a physically rigorous model of the data-acquisition system and discretization of the permittivity map of the target being imaged \cite{Brown2019} makes these techniques computationally expensive. Even with the use of calibration techniques that lessen demands on accurate models of the data-acquisition system, these optimization-based inversion methods require many iterations of a forward solver before converging. In addition, the non-uniqueness of the underlying ISCP means that they don't always converge to the true permittivity map for the OI/ROI.

In recent years, machine learning—particularly deep learning—has gained significant traction for MWI~\cite{chen2020review}. Deep learning methods have been applied across nearly all stages of the imaging pipeline, including calibration, post-processing, image enhancement, and inverse problem solving \cite{khoshdelThesis}. Among these tasks, addressing the associated ISCP is clearly the most challenging but can have the greatest impact. Supervised learning approaches attempt to bypass iterative optimization by directly mapping scattered-field measurements to target reconstructions using large datasets of measurement/target pairs~\cite{ongie2020deep,Khoshdel2020, kh2023multiPedram}. Studies have shown that deep networks can achieve high-quality reconstructions and even enable real-time imaging~\cite{kh2023multiPedram}, particularly when combined with physics-informed pre-processing such as the Born approximation or electromagnetic backpropagation~\cite{shao2020microwave, Ambrosanio2022, zumbo2024miphduo}.

A critical limitation of both traditional inversion techniques and deterministic Machine-Learning (ML) frameworks is that, by definition, they output a single reconstruction for a given input, neglecting the fundamental non-uniqueness of the electromagnetic inverse problem.  In reality, multiple plausible target permittivity maps may correspond to the same measurement data \cite{devaney2012mathematical, colton1998inverse}. In deterministic machine-learning frameworks this one-to-one  map is ultimately created by the training data set and thereby limits the trained ML model's generalizability to making predictions on scattered-field data corresponding to unseen targets that are too far away from the training data. That is, it is widely recognized that such ML models are highly sensitive to ``domain gaps'' \cite{khoshdel2021multibranch}. 

In addition to the deterministic nature of many supervised ML models, there is the related issue that these approaches require large datasets for training. To address this, researchers have employed generative models such as Generative Adversarial Networks (GANs) and diffusion models to create large labeled datasets for subsequent use in supervised learning~\cite{khoshdel2024ddpm}. For instance, Shao \textit{et al.}~\cite{Shao2022DeepLearning} combined a GAN with a deep neural network to approximate electromagnetic scattering in microwave breast imaging, thereby generating additional training data.

Although synthetic datasets are convenient, they often fail to capture practical complexities such as antenna coupling, calibration errors, and hardware imperfections\cite{martin2023cycle}. Therefore, although large synthetically generated datasets have indeed been used for training, resulting ML models typically require either augmenting the training set with experimental data or “calibrating out’’ the experimental setup to make the data more representative of synthetic conditions. Otherwise, networks trained exclusively on synthetic data often struggle to generalize to experimental measurements acquired using unique lab-specific systems. It should be commented that currently there does not exist a standardized MWI data-acquisition setup--most researchers have developed there own unique systems.

Thus, the gap between synthetic training data and experimental measurements remains a major barrier to deploying generalizable deep learning–based MWI systems. On the other hand, generative approaches are inherently less sensitive to domain discrepancies, as they learn the underlying data manifold rather than a deterministic mapping, thereby mitigating the impact of distributional shifts between training and testing datasets.

Most recently, researchers have leveraged the inherent non-uniqueness of generative models to tackle the non-unique nature of inverse problems. Early efforts predominantly employed GANs and VAEs. For example, Bhadra \textit{et al.} introduced an image-adaptive GAN framework that allows high-fidelity reconstruction of under-sampled MRI data and improved data consistency in ill-posed inverse imaging problems~\cite{GANSS}. Similarly, Yao \textit{et al.}~\cite{Yao2024CGANScattering} introduced a conditional GAN framework to map scattered electromagnetic fields directly to dielectric contrasts, enabling real-time image reconstruction. Despite these advancements, GAN-based approaches often suffer from training instabilities, mode collapse, and limited control over output diversity.

To overcome these limitations, diffusion and score-based generative models have emerged as more robust, theoretically grounded alternatives. These models have demonstrated state-of-the-art performance across a broad range of imaging tasks, including synthesis, reconstruction, and enhancement~\cite{Song2019Generative, Ho2020Denoising, Song2021ScoreBased}. Compared with GANs and VAEs, diffusion-based approaches exhibit stable optimization dynamics, provide well-defined likelihood formulations, and yield superior generative fidelity. Accumulating evidence indicates that they consistently surpass earlier generative paradigms in various medical imaging applications~\cite{Song2019Generative, Ho2020Denoising, SohlDickstein2015Nonequilibrium, Michigan_Inverse}.

In the domain of medical image reconstruction, Song \textit{et al.}~\cite{Michigan_Inverse} proposed a score-based generative framework capable of reconstructing medical images from partial CT and MRI measurements in an unsupervised manner, achieving strong generalization across diverse measurement processes. Nonetheless, applications of such models as inverse solvers for MWI remain limited. Recently, Bi \textit{et al.}~\cite{Bi2024DiffusionEMIS} introduced DiffusionEMIS, a diffusion-based method that iteratively refines 3-D point clouds to reconstruct scatterer geometries conditioned on measured scattered fields.

Building on these advances, we propose a conditional diffusion‑based generative model that explicitly incorporates the non‑unique nature of the microwave inverse scattering problem. Unlike deterministic supervised networks that produce a single estimate for a given set of measurements, our framework generates multiple plausible reconstructions consistent with the same data, thereby reflecting the inherent non-uniqueness of the ill-posed problem and leveraging the strengths of diffusion models to produce physically meaningful solutions. The core innovation of our approach lies in the integration of a physics-based selection mechanism, transforming the framework into a physics-informed generative system. After the diffusion model produces multiple candidate reconstructions, a forward electromagnetic solver is applied to each candidate to predict the corresponding scattered fields. The reconstruction yielding the lowest data discrepancy with respect to the measured fields is reported as the final solution. This physics-guided validation ensures that the chosen reconstruction is not only statistically plausible, based on the conditioning data, but also physically consistent with the underlying electromagnetic principles.

This integration of a forward solver marks a departure from purely data-driven approaches, forming a hybrid architecture that combines the generative flexibility of diffusion models with the physical accuracy of electromagnetic modeling. By embedding physics-based validation directly within the inference process, our method effectively bridges the gap between machine learning efficiency and the physics of electromagnetic interactions with the target, addressing one of the major limitations of deep learning-based inverse solvers.

The proposed methodology is particularly well-suited for data collected using actual MWI systems, where measurement noise, calibration errors, and model mismatches often degrade performance. By generating multiple candidate solutions and selecting the most physically consistent one, our approach achieves enhanced robustness and generalization. We validate the method using both synthetic and experimental datasets, demonstrating improved reconstruction accuracy and stability compared to deterministic baselines.

To address the challenges outlined above, the remainder of this paper is organized as follows. Section II provides an overview of the methodology and fundamental concepts behind diffusion models, including the forward and reverse processes and the use of diffusion priors for inverse problems. Section III introduces the proposed latent diffusion framework, which integrates a latent autoencoder representation, a physics-aware conditioning mechanism, and an end-to-end inversion architecture augmented by a physics-based selection strategy. Section IV describes the dataset used for training and evaluation, followed by Section V, which presents the experimental results, including reconstruction consistency on synthetic data, generalization to experimental measurements, multi-frequency inversion, and performance evaluation of model error. Finally, Section VI concludes the paper with key findings and future directions, and Section VII acknowledges supporting contributions.

\section{Methodology}

\subsection{Diffusion Models}

\paragraph*{Forward and reverse processes.}
Diffusion models aim to approximate a target data distribution by constructing a forward process that incrementally adds Gaussian noise and a reverse process that learns to remove this noise. Given a clean sample $x_{0}\sim p_{\mathrm{data}}$, the forward Markov chain $\{x_{t}\}_{t=1}^{T}$ evolves as
\[
x_{t} = \sqrt{1-\beta_{t}}\,x_{t-1} + \sqrt{\beta_{t}}\,w_{t},
\quad w_{t}\stackrel{\text{i.i.d.}}{\sim}\mathcal{N}(0,I),\quad 1\le t\le T,
\]
where the noise schedule $\{\beta_{t}\}$ governs how quickly the signal is corrupted. As $t$ increases, $x_{t}$ approaches an isotropic Gaussian.

Generation proceeds by starting from a Gaussian $x^{\mathrm{rev}}_{T}\sim\mathcal{N}(0,I)$ and iteratively denoising back to $x^{\mathrm{rev}}_{0}$:
\[
x^{\mathrm{rev}}_{T}\longrightarrow x^{\mathrm{rev}}_{T-1}\longrightarrow \cdots \longrightarrow x^{\mathrm{rev}}_{0}.
\]
Sampling algorithms for this reverse chain—deterministic or stochastic—can be interpreted as discretizations of underlying ODEs or SDEs. They rely on the \emph{score function} $s_{t}^{\star}(x) = \nabla_{x}\log p_{x_{t}}(x)$ of the intermediate distributions. Tweedie’s formula provides an explicit expression,
\[
s_{t}^{\star}(x)
= -\frac{1}{1-\alpha_{t}}
\int p_{X_{0}\mid X_{t}}(x_{0}\mid x)\,\bigl(x - \sqrt{\alpha_{t}}\,x_{0}\bigr)\,\mathrm{d}x_{0},
\]
with $\alpha_{t} := 1-\beta_{t}$ and $\alpha_{t}:=\prod_{k=1}^{t}\alpha_{k}$. In practice, one trains a neural network to approximate these score functions via score matching, enabling the learned model to generate samples from $p_{\mathrm{data}}$.

\paragraph*{Diffusion priors for inverse problems.}
The flexibility of diffusion models makes them attractive as priors for ill‑posed inverse problems. Suppose we observe data $y$ generated from an unknown signal $x^{\star}$ through a known forward operator and additive noise. A Bayesian formulation samples from the posterior
\[
p(x\mid y)\propto p_{\mathrm{prior}}(x)\,p\bigl(y\mid x\bigr),
\]
where $p_{\mathrm{prior}}(x)$ is the diffusion‑model prior and $p(y\mid x)$ is the likelihood. Sampling from this posterior combines the learned score functions with the physics of the measurement process.

\begin{figure*}[!t]
    \centering
    \begin{tabular}{cc}
        \subfloat[Training the model]{\includegraphics[width=18cm]{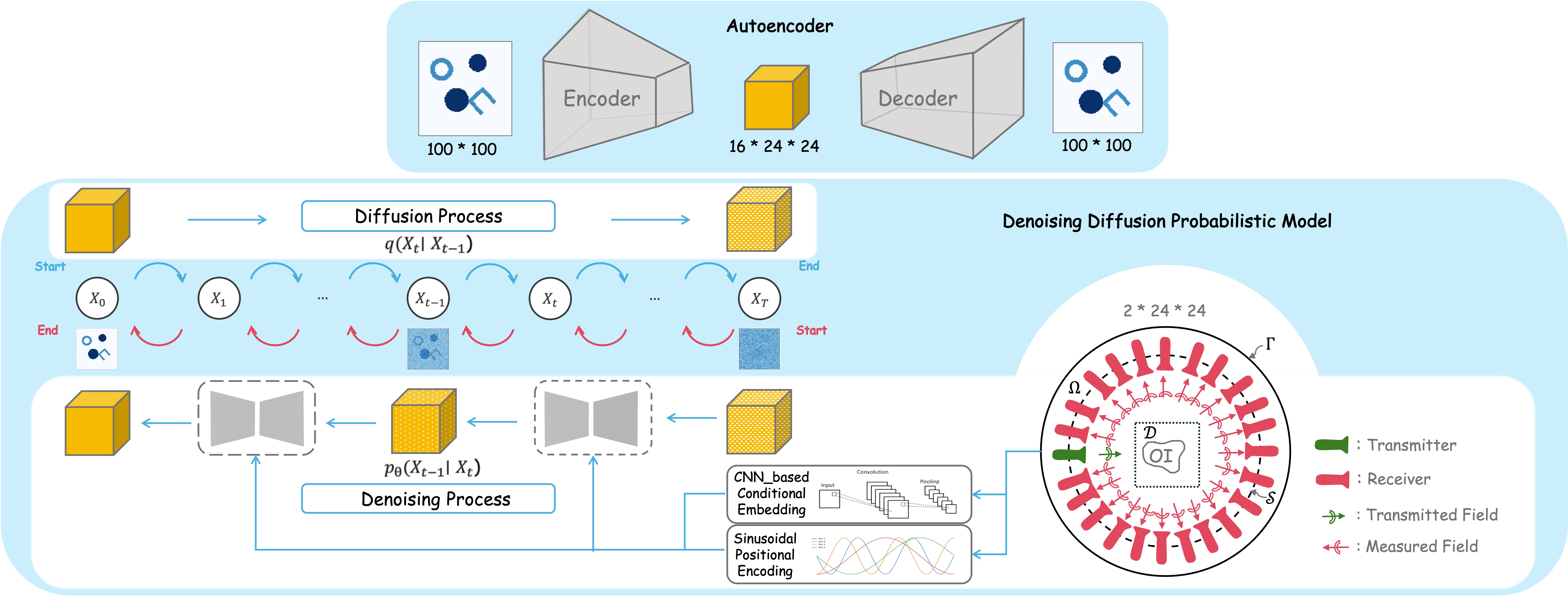}} \\
        \subfloat[Inference]{\includegraphics[width=18cm]{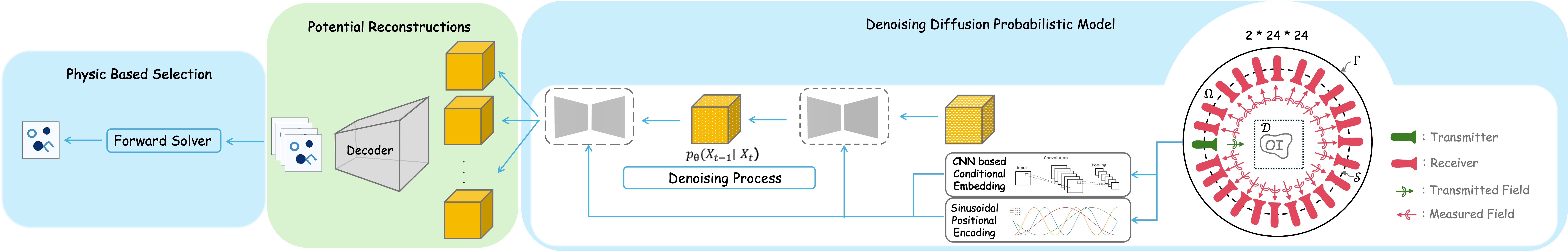}} \\
    \end{tabular}
    \caption{Overview of the proposed framework: (a) training phase of the denoising diffusion model; (b) inference phase for reconstructing the output from measured fields.}
    \label{fig:training_inference}
\end{figure*}

\subsection{Proposed Latent Diffusion Model}
We consider the electromagnetic inverse problem of reconstructing two-dimensional permittivity distributions from scattered field measurements. The scattered fields are represented as two channels, corresponding to the real and imaginary components of the electric field. To overcome the non-uniqueness of the inverse problem, we design a generative inversion framework based on a conditional diffusion model operating in a compact latent space of permittivity maps. This framework produces multiple plausible reconstructions that are driven by, \textit{i.e.}, consistent with, the measured fields, after which a post-processing step is applied to identify the most physically meaningful solution.

\subsubsection{\textbf{Latent Representation with Autoencoder}}

Directly applying a diffusion model to full-resolution permittivity maps is computationally expensive, particularly when extending this work to 3D medical imaging applications, which represent the ultimate goal of our research. To address this, we first train an autoencoder (AE) to learn a compact latent representation, following a design similar to the one used in ~\cite{LatentDiffusion_2022_CVPR}.
\begin{itemize}
    \item \textbf{Encoder (E):} compresses the permittivity grid $x$ into a latent representation $z=E(x)$ using convolution and downsampling layers.
    \item \textbf{Decoder (D):} reconstructs the grid from the latent code as $\tilde{x} = D(z) = D(E(x))$ using upsampling operations.
\end{itemize}

The AE is trained with a composite loss function:
\begin{itemize}
    \item Reconstruction loss (pixelwise error) ensures quantitative fidelity to the input.
    \item  Perceptual loss~\cite{PerceptualLosses}, computed from intermediate features of a pretrained VGG16 network  ~\cite{VGG}, encourages preservation of edges and structural features.
    \item  Adversarial loss ~\cite{goodfellow2014generative} penalizes overly smooth reconstructions and promotes realistic textural detail.
\end{itemize}

The relative weights of these terms are tuned to balance numerical accuracy with perceptual quality. By compressing a 100×100 grid into a 16×24×24 latent vector, the AE reduces the computational cost of the subsequent diffusion process.

\subsubsection{\textbf{Physics‑Aware Conditioning Mechanism}}

The learned latent space serves as the domain for our diffusion model. The model is conditioned on  measured scattered fields collected in the scenario depicted in Fig. xxxx where a dielectric target is located at the center of a ring of transmitter and receivers. The transmitters and receivers are located at equidistant points on a circle of radius XXX m surrounding the target. A complete descriprion can be found in \cite{cathers2025im}. This condition ensures that the generated permittivity maps are physically consistent with the observations. The data is composed of a 24×24 scattered-field matrix of complex-valued data representing the real and imaginary parts of the received frequency-domain phasor. Data consists of up to five frequencies collected at 3.0, 3.5, 4.0, 4.5, and 5.0 GHz. 

\begin{itemize}
	\item \textbf{Forward diffusion:} Gaussian noise is added to latent vectors over a predefined schedule.
	\item \textbf{Reverse denoising:} A U-Net–based denoiser is trained to iteratively predict and remove this noise. The denoiser is conditioned on the scattered fields via cross-attention layers.
	\item \textbf{Conditioning mechanism:} In the proposed framework, the scattered electromagnetic fields are initially processed through convolutional layers to project them into a feature space. Because electromagnetic inverse problems depend on the spatial arrangement of transmitters, receivers, and objects, it is vital to retain positional information—something that standard convolution and pooling layers tend to lose. To address this, we augment the 24×24 scattered-field inputs with sinusoidal positional encodings that explicitly encode transmitters coordinates. Injecting these encodings into the network provides an absolute reference frame, enabling the model to distinguish sensor locations and object orientations. This spatial awareness is especially important for microwave imaging, where the geometry of the scene strongly influences the observed scattering.
    These features are then injected back into the diffusion model’s denoiser at every step: they are concatenated with intermediate feature maps and linked via cross-attention blocks so that the denoising operation remains conditioned on the measured electromagnetic response. By coupling the generative model to the observed physics in this way, we encourage it to produce reconstructions that are physically consistent with the actual scattered fields.

	\item \textbf{Architecture:} The U-Net consists of downsampling and upsampling blocks with intermediate attention ~\cite{U-Net}. Time embeddings indicate the current noise level, while projected scattered field features provide physical guidance.
\end{itemize}
The denoiser is trained using a Mean-Squared-Error (MSE) between the true and predicted noise. This objective allows the network to implicitly learn the conditional distribution of latent permittivity representations.

\subsubsection{\textbf{End-to-End Inversion Framework}}

The full reconstruction pipeline proceeds as follows:
\begin{itemize}
	\item \textbf{Encoding:} The AE encoder maps the permittivity grid to a latent representation.
	\item \textbf{Diffusion training:} The conditional diffusion model learns to denoise noisy latent codes, conditioned on scattered fields.
	\item \textbf{Sampling:} For unseen measurements, the diffusion model generates latent permittivity representations driven by the observed fields.
	\item \textbf{Decoding:} The AE decoder reconstructs the full-resolution permittivity grid.
\end{itemize}
This framework, illustrated in Fig.~\ref{fig:training_inference}, combines the representational efficiency of autoencoders with the generative power of diffusion models, resulting in a distribution of generated high-quality reconstructions of permittivity maps from  electromagnetic data. 

\subsubsection{\textbf{Physic Based Selection}}

The conditional diffusion model produces a set of reconstructions for \textbf{one} scattered field data that are broadly consistent with the measured scattered fields, mirroring the non-uniqueness of the ill-posed inverse problem. That is, after training, the diffusion model is used to make several inferences for each measurement, generating a set of candidate permittivity maps. To choose between the reconstructed candidates, we introduce a post-processing stage designed to select the most physically meaningful reconstruction.  Specifically, each  candidate is  passed through a forward solver to compute the scattered fields it would produce, illustrated in part b of Fig.~\ref{fig:training_inference}. We compute the MSE between the simulated and measured scattered fields, choosing the candidate permittivity map that produces the minimum scattered-field MSE, providing a global measure of accuracy. By relying exclusively on scattered-field MSE we do not use any prior information regarding the actual permittivity map.

In summary, the conditional diffusion model ensures that the candidate permittivity maps appear physically reasonable, and the forward-solver selection process ensures that the corresponding scattered fields  closely match the experimental measurements. This paradigm opens the door for post-processing selection procedures based on different criteria and/or other available prior information, \textit{e.g}, smoothness of the permitivity map.

\subsection{Dataset}
\begin{figure}
  \centering
  \includegraphics[width=0.9\columnwidth]{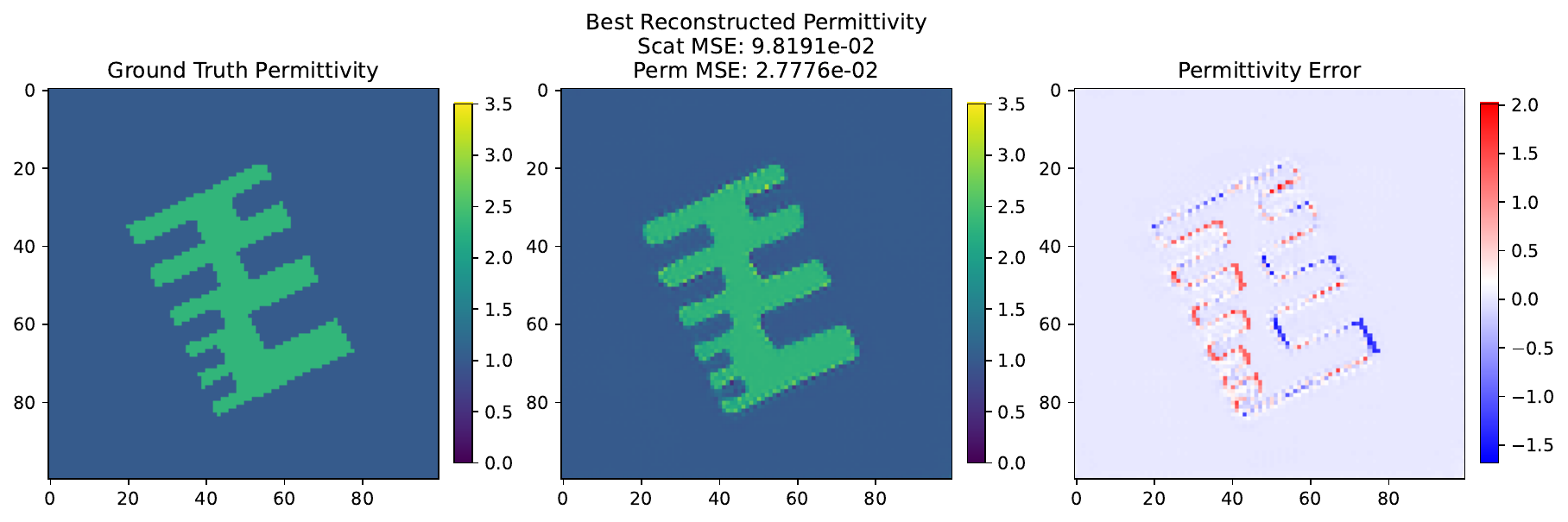}
  \hfill
  \includegraphics[width=0.9\columnwidth]{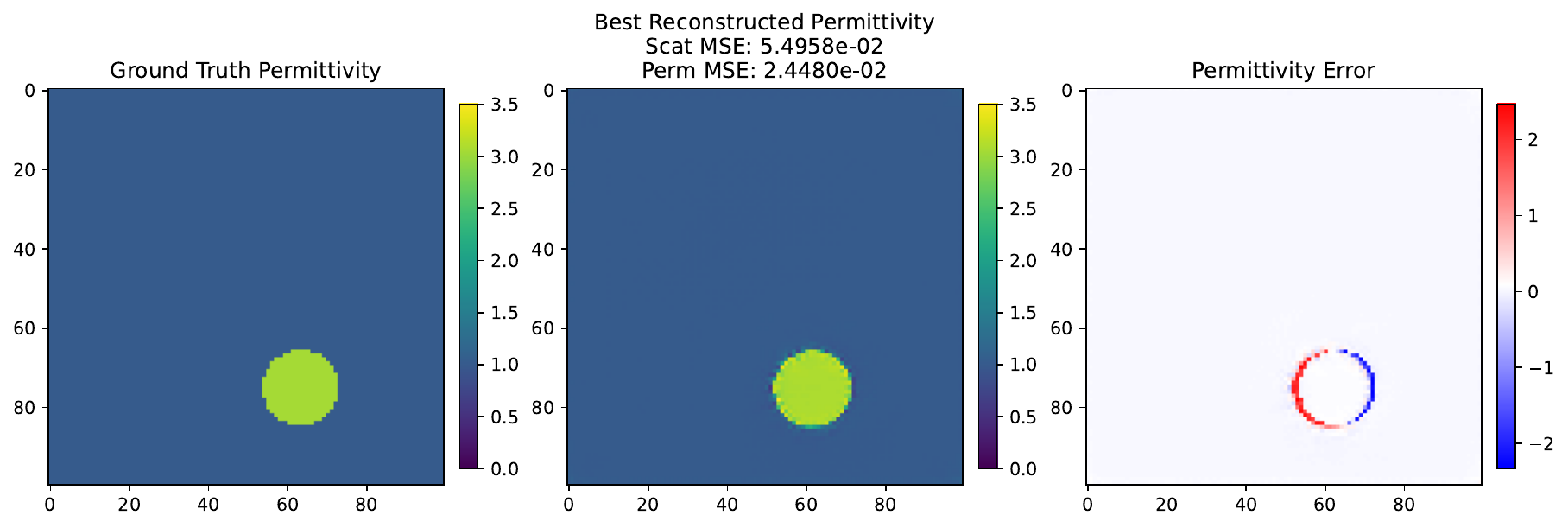}\\[1ex]
  \includegraphics[width=0.9\columnwidth]{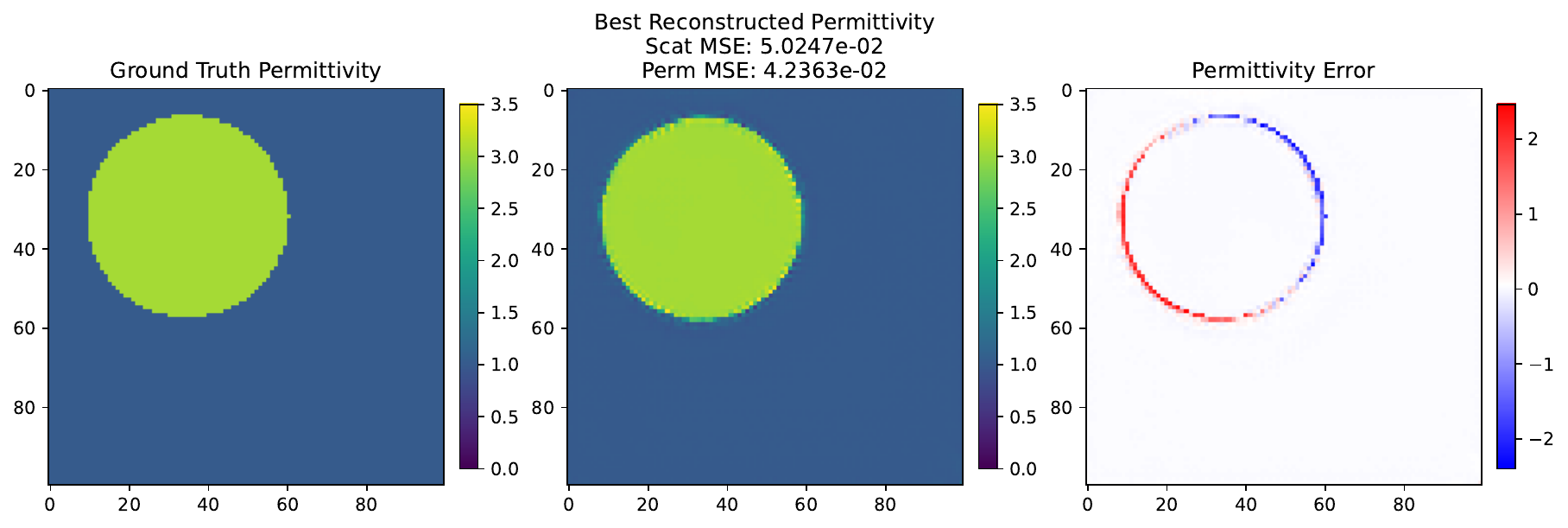}
  
  \caption{Evaluation of the proposed model using single frequency synthetic measurements for three representative samples from each category. The best reconstruction is selected from 100 generated candidates using the proposed physics-guided selection mechanism.}
  \label{fig:synthetic}
\end{figure}

Two datasets are used in this study. The first dataset, DataSet1, is the publicly available benchmark presented by Cathers \textit{et al.}\cite{Cathers2025}, which includes both synthetic and experimental measurements. It contains scattering data for Nylon-66 cylinders with diameters of 3.8 cm and 10.2 cm (reported permittivity: $\epsilon_r = 3.03 - j0.03$) and a complex-shaped object referred to as the E-phantom, machined from an ultra-high molecular weight polyethylene (UHMWP) block (reported permittivity: $\epsilon_r = 2.3$\cite{Bigelow1999}). The cylinders were translated spatially to produce multiple configurations within the imaging domain, while the E-phantom was both translated and rotated to generate a diverse set of measurement data. For each target, the dataset provides both synthetic scattered fields, computed using a 2D scalar Method-of-Moments (MoM) forward solver based on Richmond’s method~\cite{Richmond} on a 100 × 100 permittivity grid, and calibrated experimental measurements acquired under comparable conditions~\cite{cathers2024description, cathers2025im}. Representative examples of the synthetic and experimental data for each target are illustrated in first column of Figure~\ref{fig:synthetic}.

\begin{figure*}[!b]
    \centering
    \includegraphics[width=0.95\linewidth]{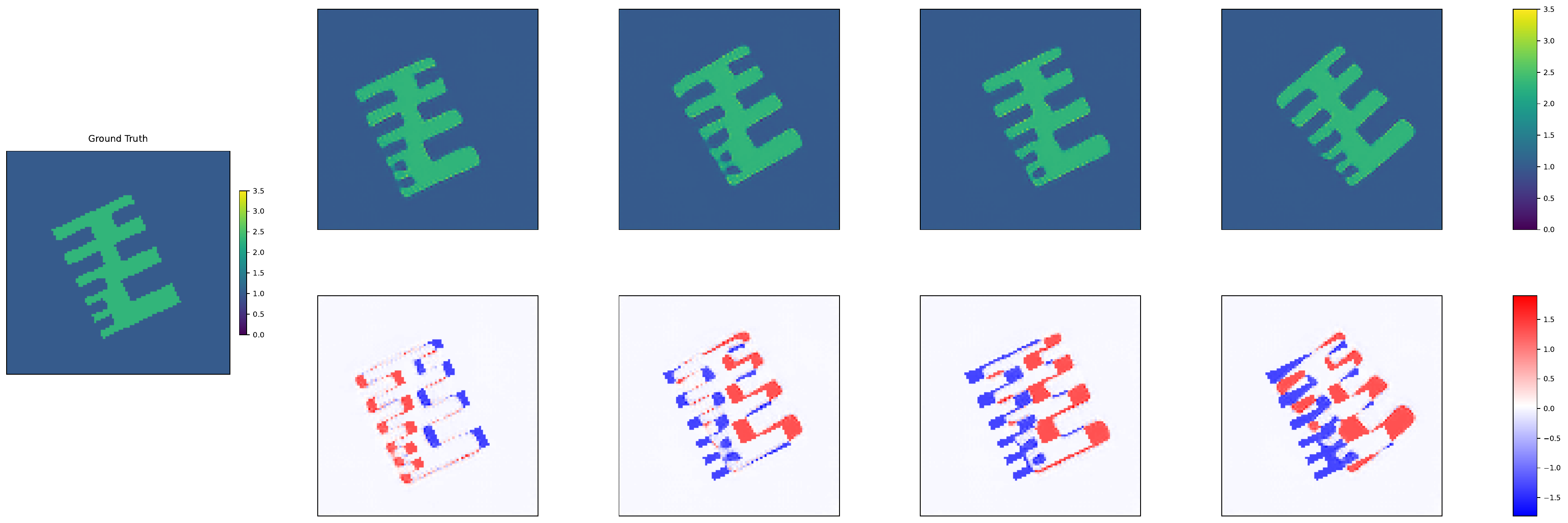}
   \caption{Four candidate reconstructions of the same scattered-field data at 5 Ghz generated from different random noise initializations, illustrating the generative and stochastic nature of the diffusion-based model.}
    \label{fig:5_samples}
\end{figure*}
Although DataSet1, obtained from \cite{cathers2024description, Cathers2025}, provides valuable benchmarking data, it includes a limited number targets displaying a restricted class of scattering features.  This limits one's ability to properly investigate and evaluate an machine-learning model's robustness and generalizability as an inverse solver. To address this limitation and assess the robustness of the proposed inverse solver, we created a new dataset comprising 2000 samples of dielectric targets (DataSet2). The scattered fields were generated under ideal 2D point-source illumination (3D line sources) using the same 2D scalar MoM forward solver on a 100 × 100 permittivity grid. This dataset introduces composite dielectric targets made up of random configurations of basic \textit{canonical} dielectric shapes. The canonical dielectric shapes are circles of varying sizes, hollow circles, and U-shapes of different dimensions. Although the permittivity of each is uniform across the shape, one of two possible values of permittivity are chosen, either $\varepsilon_r = 2.3$ or $\varepsilon_r = 3.03$, for any particular shape. Each canonical shape introduces challenging scattering features of its own, \textit{e.g.}, the hollow circles introduces the feature of penetrability of energy, and multiple scattering from within the target. The U-shaped targets introduce direction-based, or anisotropic, scattering. Both have stronger frequency-dependent scattering features than the simple solid circles of DataSet1. Although the E-phantom of DataSet1 is approximately a combination of overlapping canonical U-shapes, in the E-phantom these are fixed with respect to their relative positions.

The positions, sizes, and orientations of the canonical shapes were randomly varied to create diverse spatial configurations. In addition, each composite target was created using up to four canonical shapes positioned at random within the grid. This produces random \textit{overlapping} intersections of the shapes, further increasing the overall geometric complexity of the final composite scatterer. Some representative composite targets in DataSet2 are depicted in Figs. \ref{fig:AE}, \ref{fig:dataset2_signle} and \ref{fig:dataset2_multi}. These two datasets enable a comprehensive variation of scattered fields allowing us to evaluate the performance of the proposed ML model.

\section{Results}
\subsection{Reconstruction Consistency on Synthetic Data}
The latent diffusion model was first trained using only synthetic data from DataSet1 and subsequently evaluated also using only unseen synthetic samples that differed in position and orientation from those in the training set. Figure~\ref{fig:5_samples} presents four independent reconstructions conditioned on scattered-field data at 5 GHz from a single instance of the E-phantom. Each of the four candidate solutions shown  was generated from a distinct random noise initialization of the diffusion model. As illustrated, the model consistently reproduces the target permittivity distribution with minimal reconstruction error, demonstrating its stability and robustness to stochastic variation in the sampling process.

To identify the best reconstruction, the diffusion model was used to generated 100 candidate permittivity maps, each evaluated using a forward electromagnetic solver to compute the corresponding scattered fields. It should be noted that the choice of 100 reconstructions is arbitrary and can be adjusted according to the time or computational constraints of each application. The reconstruction that was selected as the final result was the one yielding the lowest mean squared error (MSE) between the predicted and ground-truth scattered fields. As illustrated in Figure ~\ref{fig:synthetic} , the selected reconstruction not only achieves a low MSE in the scattered field domain but also exhibits a low MSE in the permittivity distribution, indicating that the reconstructed permittivity grid closely matches the ground truth in both material properties and resulting electromagnetic behavior.The corresponding MSE values for this part can be found in Table I, row 3.

\subsection{Generalization to Experimental Data} 
\begin{figure}
  \centering
  \includegraphics[width=0.9\columnwidth]{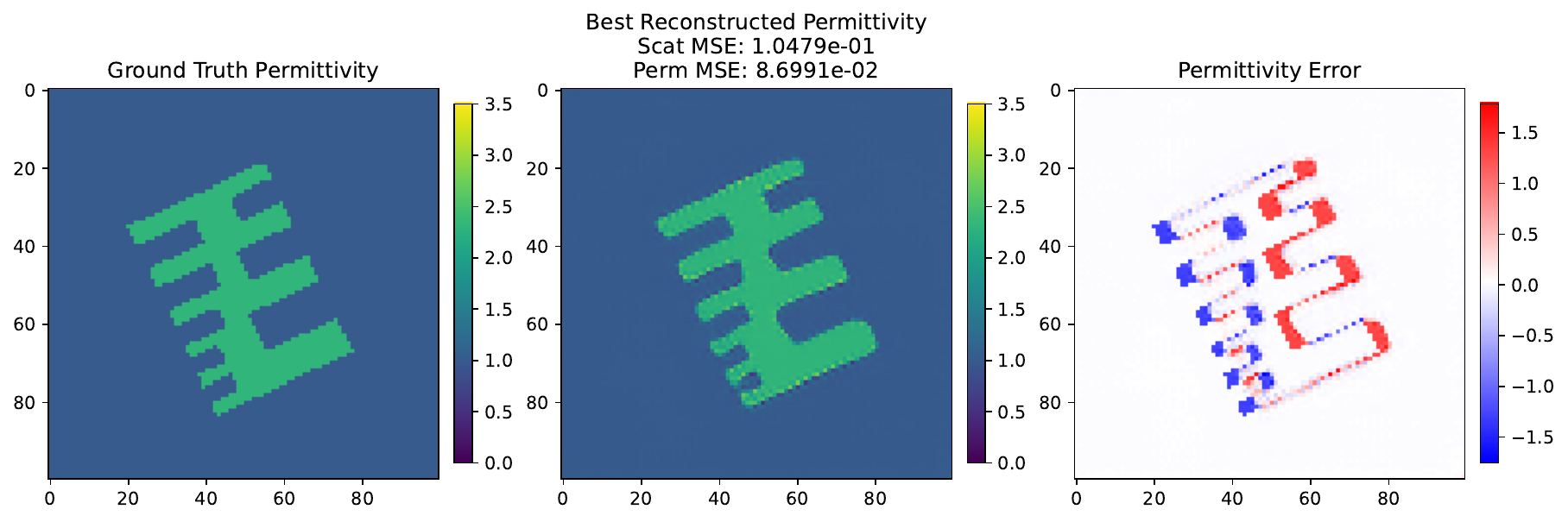}
  \hfill
  \includegraphics[width=0.9\columnwidth]{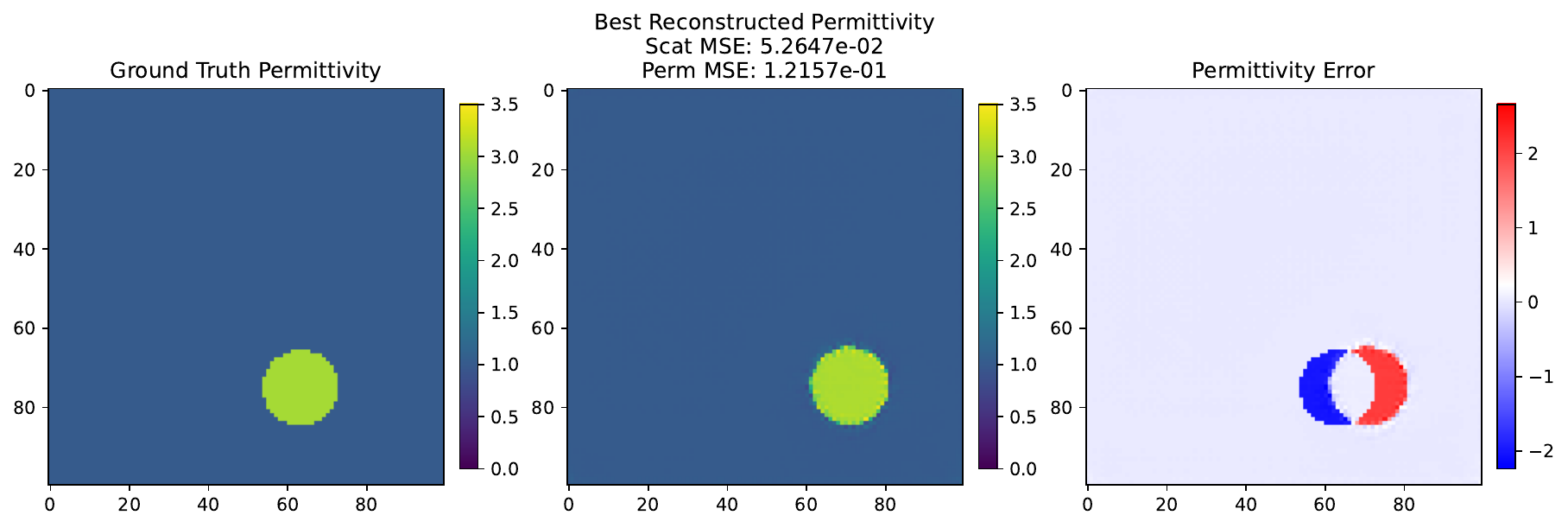}\\[1ex]
  \includegraphics[width=0.9\columnwidth]{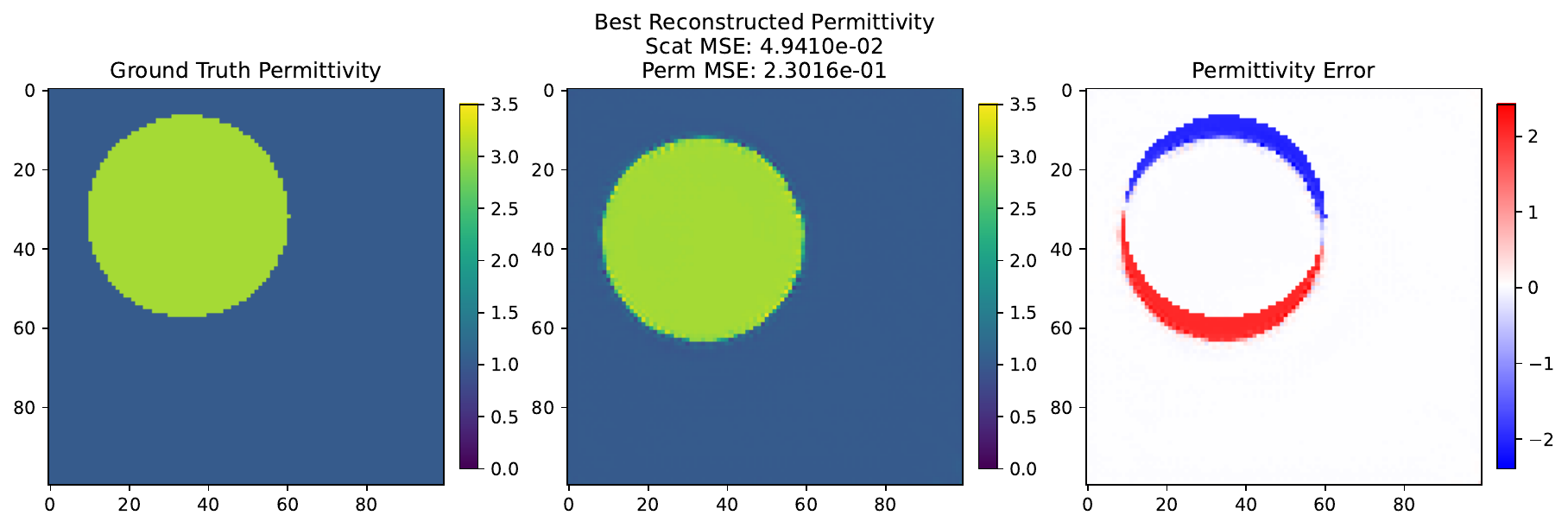}
  
\caption{Evaluation of the proposed model on single-frequency experimental measurements, where the model was trained exclusively on synthetic data. The best reconstruction is selected from 100 generated candidates using the proposed physics-guided selection mechanism.}
  \label{fig:experimental}
\end{figure}

To assess the proposed model's ability to generalize beyond synthetic training data, the trained model was subsequently evaluated on experimental scattered-field measurements from DataSet1. As shown in Figure~\ref{fig:experimental}, the model achieved promising reconstruction performance, successfully recovering key structural features from previously unseen experimental samples. Although the reconstructed permittivity maps were less accurate than those obtained when testing synthetic data, the results demonstrate the model’s capacity to adapt to experimental measurements, even though it was trained solely on synthetic data.

\begin{figure}
  \centering
  \includegraphics[width=0.9\columnwidth]{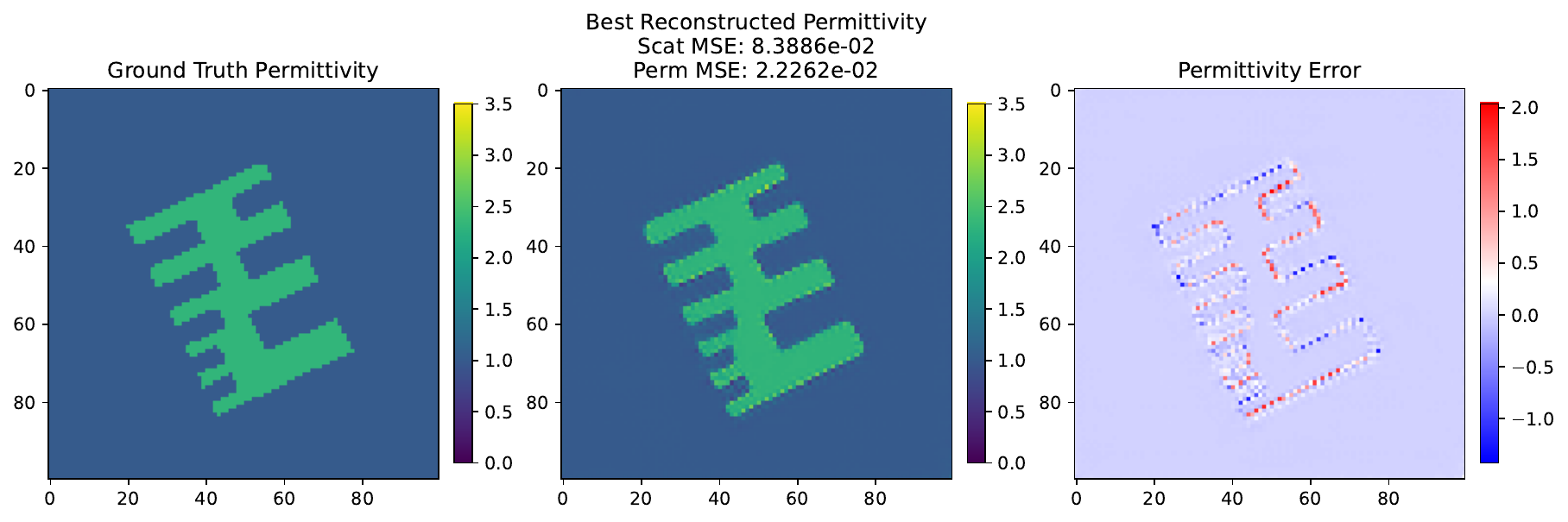}
  \hfill
  \includegraphics[width=0.9\columnwidth]{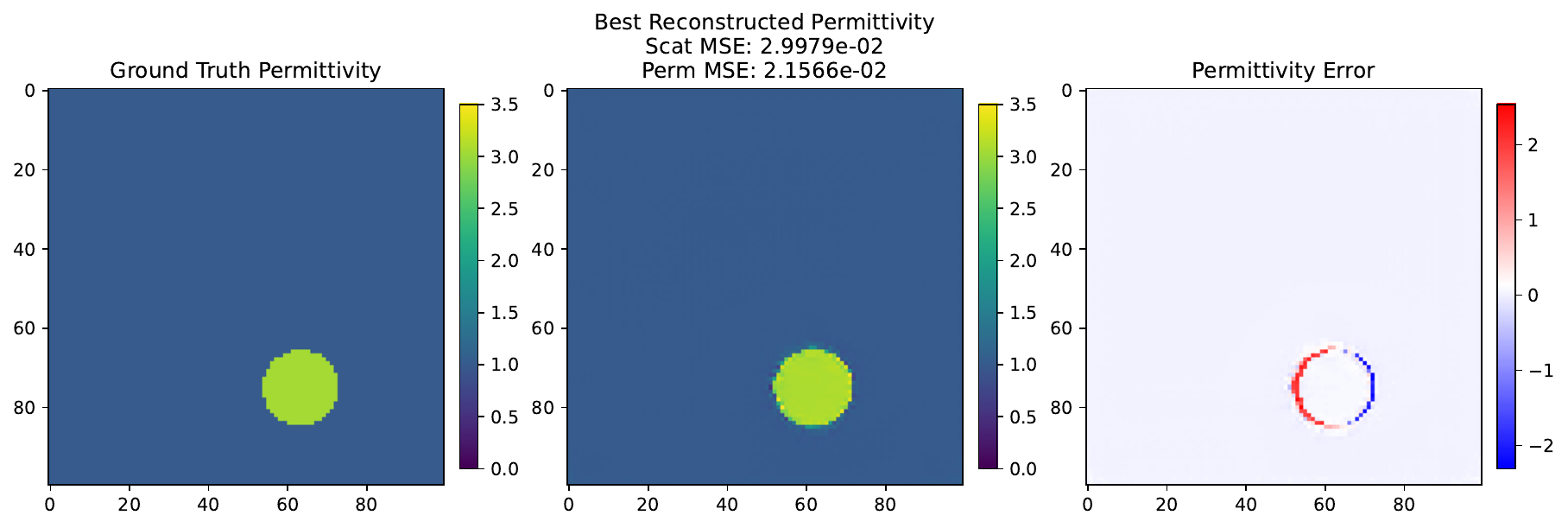}\\[1ex]
  \includegraphics[width=0.9\columnwidth]{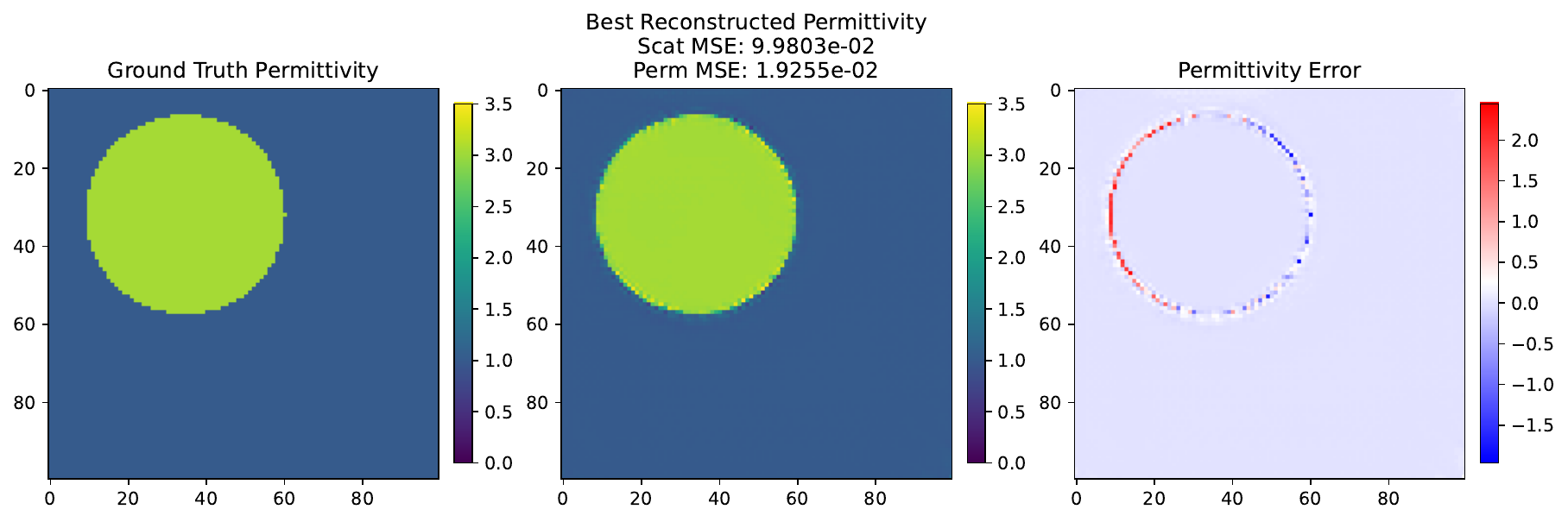}
  
  \caption{Evaluation of the proposed model on experimental measurements, where the model was trained exclusively on using multi-frequency synthetic  data generated using data from five distinct frequencies—3.0, 3.5, 4.0, 4.5, and 5.0 GHz. The best reconstruction is selected from 100 generated candidates using the proposed physics-guided selection mechanism.}
  \label{fig:five}
\end{figure}

This generalization result is notable given the significant differences between synthetic and experimental measurements, particularly for the near-field electromagnetic imaging system from which the experimental data was obtained \cite{cathers2024description}. Near-field systems, especially those utilizing co-resident antennas, exhibit field behaviors that can differ significantly from the fields generated using, necessarily approximate, computational models. For example, the synthetic training data was generated using the incident fields of idealized point sources (line-soures) and the received fields were simply taken to be scattered fields at precise point-locations. In contrast, the experimental data was collected as microwave scattering parameters (S-parameters) at the antenna ports. An antenna effectively integrates the spatially varying fields  across its aperture.  In addition,  the mutual-coupling with nonactive antennas is ignored in the synthetically generated data. Despite these discrepancies, our ML model was able to deliver strong reconstruction performance. Of course, this is partly due to the scattered-field calibration procedure that was implemented \cite{ostadrahimi2011analysis, cathers2024description}. Incorporating the calibration  procedure into a machine learning model of its own has been attempted in the past and is a research subject of the greatest importance.

\begin{table}[h!]
\centering
\renewcommand{\arraystretch}{1.5}
\caption{Quantitative comparison of average MSE values between reconstructed and ground-truth permittivity maps ($\text{MSE}_{\text{image}}$) and average MSE between true scattered-field data and corresponding scattered-field data  $(\text{MSE}_{\text{data}})$ for various training and testing conditions. }
\begin{tabular}{|p{5cm}|p{1.2cm}|p{1.2cm}|}
\hline
\textbf{Test Case} & \textbf{$\text{MSE}_{\text{image}}$} & \textbf{$\text{MSE}_{\text{data}}$} \\\hline

\noalign{\hrule height 1pt}

\multicolumn{3}{|>{\columncolor{gray!15}}l|}{\rule[0pt]{0pt}{1.2em}\textbf{Dataset 1}}\\[-0.5ex]
\noalign{\hrule height 1pt}

CNN based Model Train on Synthetic and Test on Synthetic~\cite{Cathers2025} & 0.163 & NA \\\hline
CNN based Model Train on both Synthetic and Experimental / Test Experimental~\cite{Cathers2025} & 0.089 & NA \\\hline
{Proposed model Train on Synthetic / Test  on Synthetic (Single Frequency)} & 0.0590 & 0.0848 \\\hline
Proposed model Train on Synthetic / Test Experimental (Single Frequency) & 0.0905 & 0.0869 \\\hline
Proposed model Train on Synthetic / Test Synthetic (Multi-Frequency) & 0.0334 & 0.0669 \\\hline

\noalign{\hrule height 1pt}
\multicolumn{3}{|>{\columncolor{gray!15}}l|}{\rule[0pt]{0pt}{1.2em}\textbf{Dataset 2}}\\[-0.5ex]
\noalign{\hrule height 1pt}

Train Synthetic / Test Synthetic (Single-Fre) & 0.0988 & 0.0735 \\\hline
Train Synthetic / Test Synthetic (Multi-Fre) & 0.0399 & 0.0679 \\\hline

\end{tabular}
\label{tab:mse_comparison}
\end{table}

\subsection{Multi-Frequency Inversion and Dataset Enhancement}
Results shown so far were obtained by conditioning the ML model on single frequency scattered-field data. It is well-known that using data from multiple frequencies, either simultaneously or via frequency-hopping,  can enhance reconstruction performance especially for non-dispersive targets~\cite{kaye2019improvement, Asefi2019}. To investigate how incorporating multiple frequencies improves the ML model’s ability to learn the relationship between the scattered-field data and the corresponding permittivity distribution, we conditioned the diffusion model on scattered-field measurements acquired at five distinct frequencies—3.0, 3.5, 4.0, 4.5, and 5.0 GHz—instead of a single frequency.

In addition, as previously mentioned, DataSet1 includes a limited range of object types—specifically, solid circularly-cylindrical targets and the E-phantom. Each sample in DataSet1 consists of a single object, albeit represented across multiple positions and orientations within the imaging grid.  To ensure that the ML model was not over-fitted to a narrow set of object geometries—a limitation commonly observed in prior studies—we developed the more diverse synthetic dataset, Dataset2. It provides  a more challenging evaluation of the model’s generalization ability.  

The model was re-trained using this newly generated dataset and tested on previously unseen samples. Representative reconstructions and their corresponding ground-truth targets are shown in Fig.~\ref{fig:dataset2_signle}. As expected, the inverse problem associated with this more complex, multi-object dataset presents increased difficulty compared to Dataset~1. Nevertheless, the model successfully reconstructed the underlying permittivity distributions, effectively capturing both the global structure and spatial organization of the objects.  

To further enhance the model’s capacity to learn the mapping between scattered-field measurements and corresponding permittivity distributions, we investigated the integration of multi-frequency data. Specifically, scattered-field measurements were acquired at five frequencies: 3.0, 3.5, 4.0, 4.5, and 5.0~GHz. Figure~\ref{fig:dataset2_multi} illustrates the model’s performance when conditioned on multi-frequency data, demonstrating improved reconstruction consistency and sharper structural recovery across test samples.  

The results indicate that while the single-frequency model occasionally struggles to accurately reconstruct the shapes of closely spaced or sharp-edged objects—sometimes confusing U-shaped targets with hollow circles (see rows~2 and~4 of Fig.~\ref{fig:dataset2_signle})—the multi-frequency model performs substantially better. By leveraging additional frequency-domain information, it accurately resolves fine structural details and preserves object boundaries even in complex spatial configurations. 

This improvement is quantitatively reflected in the last two rows of Table~\ref{tab:mse_comparison}, where the average $\text{MSE}_{\text{image}}$ values for the multi-frequency case decrease to approximately one-third of the single-frequency error. The averages were computed over 10 test cases, with each case reconstructed 100 times to select the best reconstruction based on the $\text{MSE}_{\text{data}}$.

These findings highlight the potential of incorporating richer physical information—beyond frequency diversity—to improve data representation and model robustness. Future research will explore additional labeled datasets encompassing a more comprehensive  set of composite scattering targets to further enhance the model’s capacity for accurate and stable inversion in challenging electromagnetic imaging scenarios.

While the qualitative evaluation demonstrates substantial improvements over both traditional inverse solvers such as Contrast Source Inversion (CSI) and previously reported state-of-the-art models—specifically, the reconstructed images presented in Figures 11–14 of \cite{Cathers2025}—a quantitative assessment was also conducted to provide a more comprehensive performance analysis. This evaluation includes comparisons with the state-of-the-art CNN-based supervised inverse solvers previously reported in the literature \cite{Cathers2025} and the proposed generative diffusion-based approach. Specifically, the mean squared error (MSE) was evaluated in both the scattered-field domain ($\text{MSE}_{\text{data}}$) and the image domain ($\text{MSE}_{\text{image}}$), with averages computed over 10 test samples. The results, summarized in Table~\ref{tab:mse_comparison}, clearly indicate that the proposed method outperforms state-of-the-art inverse solvers, achieves notable performance gains with multi-frequency data, and exhibits enhanced generalization capability when evaluated on Dataset~2.

\subsection{Performance Evaluation of ML Model's Error}

As the architecture of the designed ML model is quite complicated it is important to carefully analyze where the inference errors ar occuring. This analysis revealed that the reconstruction error originates not only from the diffusion component but also from the autoencoder (AE) reconstruction stage. The AE-induced error was found to be relatively minor when the model was trained on the simpler DataSet1 containing a limited number of object types, but became more pronounced when trained on the more diverse Dataset2. Table~\ref{tab:ae_comparison} presents a quantitative comparison of AE performance across both datasets, reporting the average MSE computed over 100 random samples. As shown, the reconstruction error increases for the more complex dataset, reflecting the greater difficulty of compressing and reconstructing diverse geometrical structures.  As the diffusion model works on the latent space of the AE, it is inevitable that errors in compressing the permittivity maps into the latent space would degrade the overall performance.

\begin{figure}
  \centering
  \includegraphics[width=0.9\columnwidth]{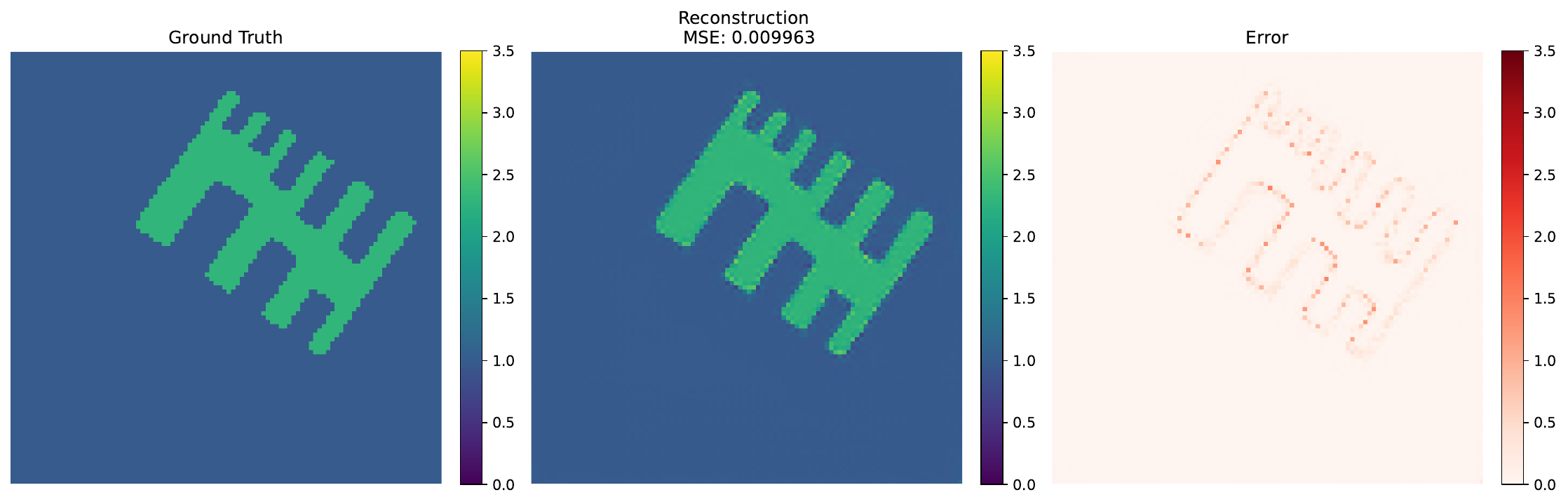}
  \hfill
  \includegraphics[width=0.9\columnwidth]{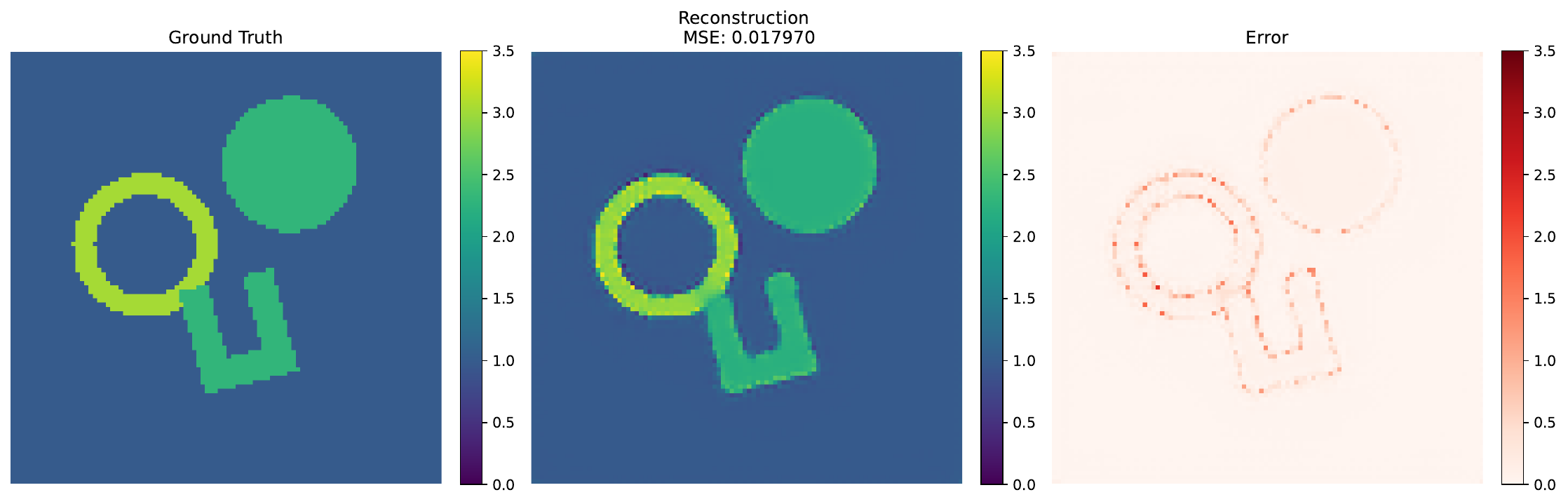}\\[1ex]
  
  \caption{Comparison of autoencoder reconstructions for DataSet1 and DataSet2.}
  \label{fig:AE}
\end{figure}

\begin{figure}
  \centering
  \includegraphics[width=0.9\columnwidth]{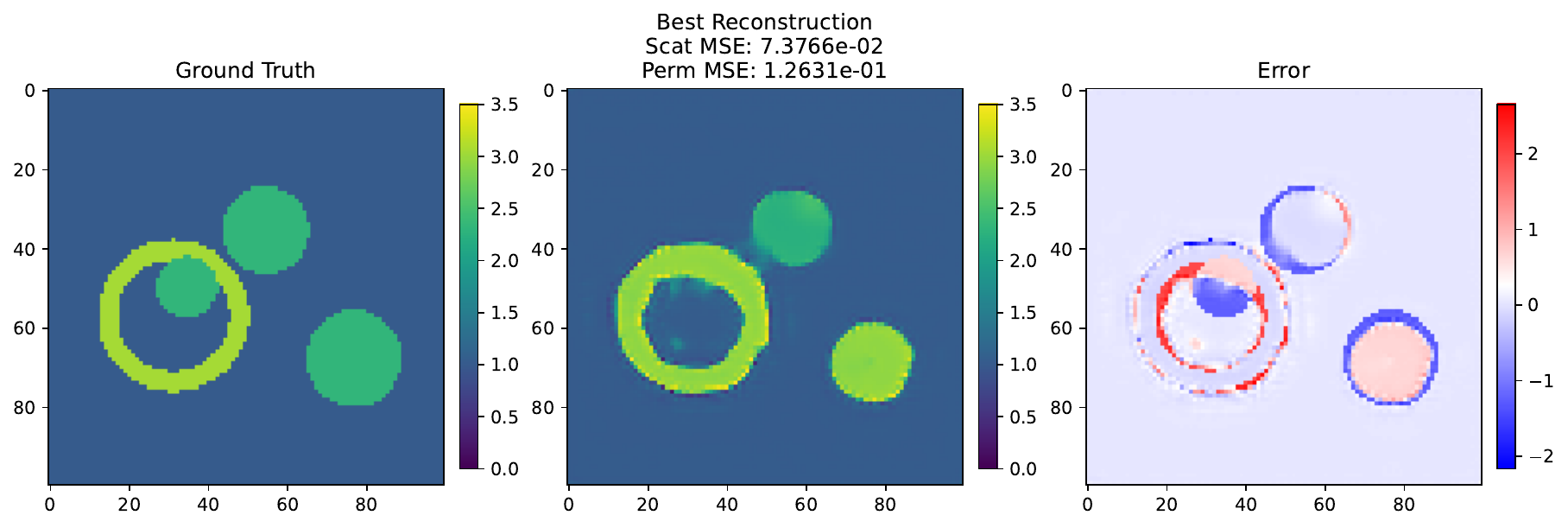}
  \hfill
  \includegraphics[width=0.9\columnwidth]{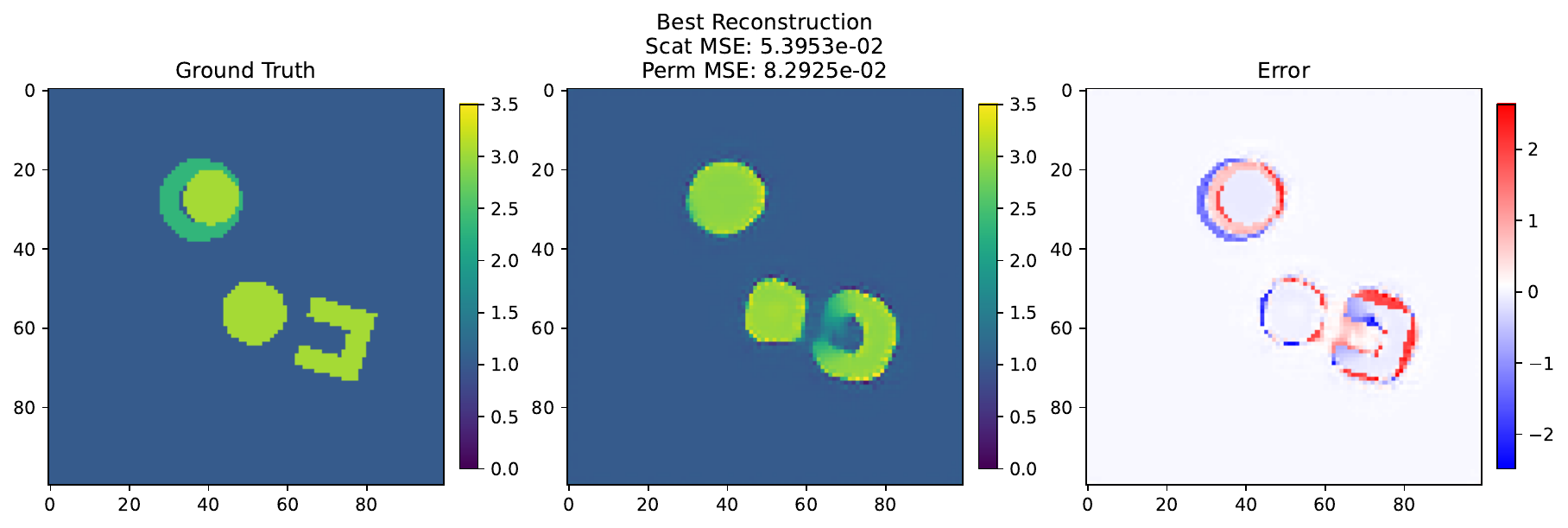}\\[1ex]
  \includegraphics[width=0.9\columnwidth]{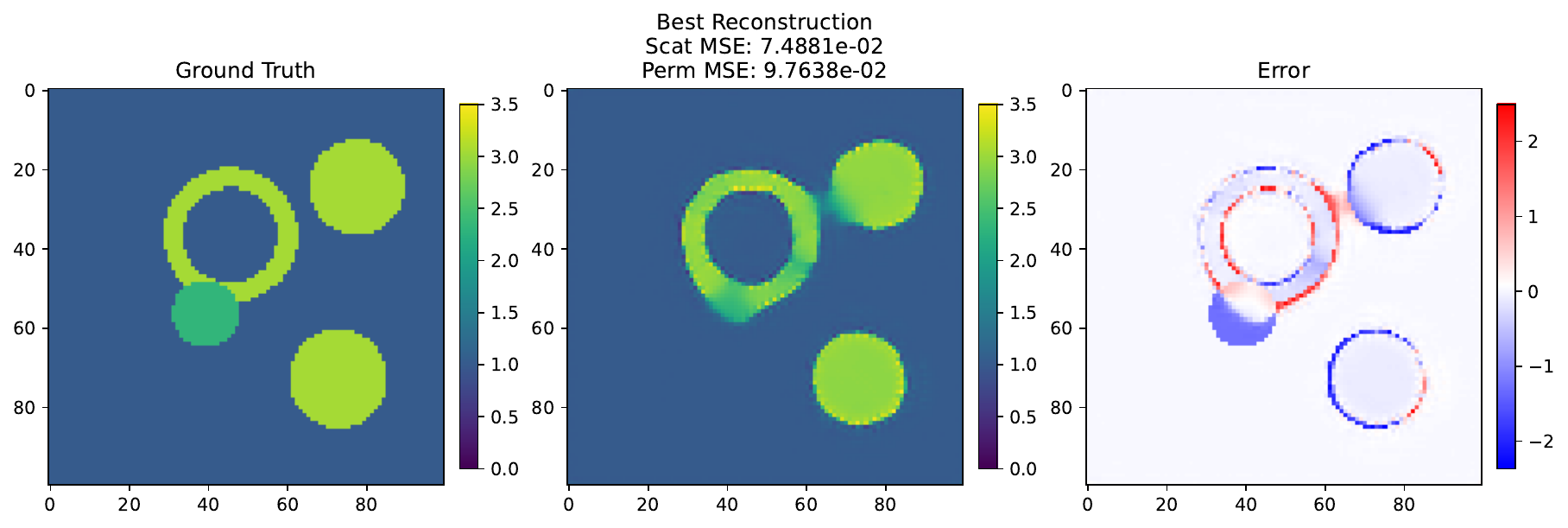}

    \includegraphics[width=0.9\columnwidth]{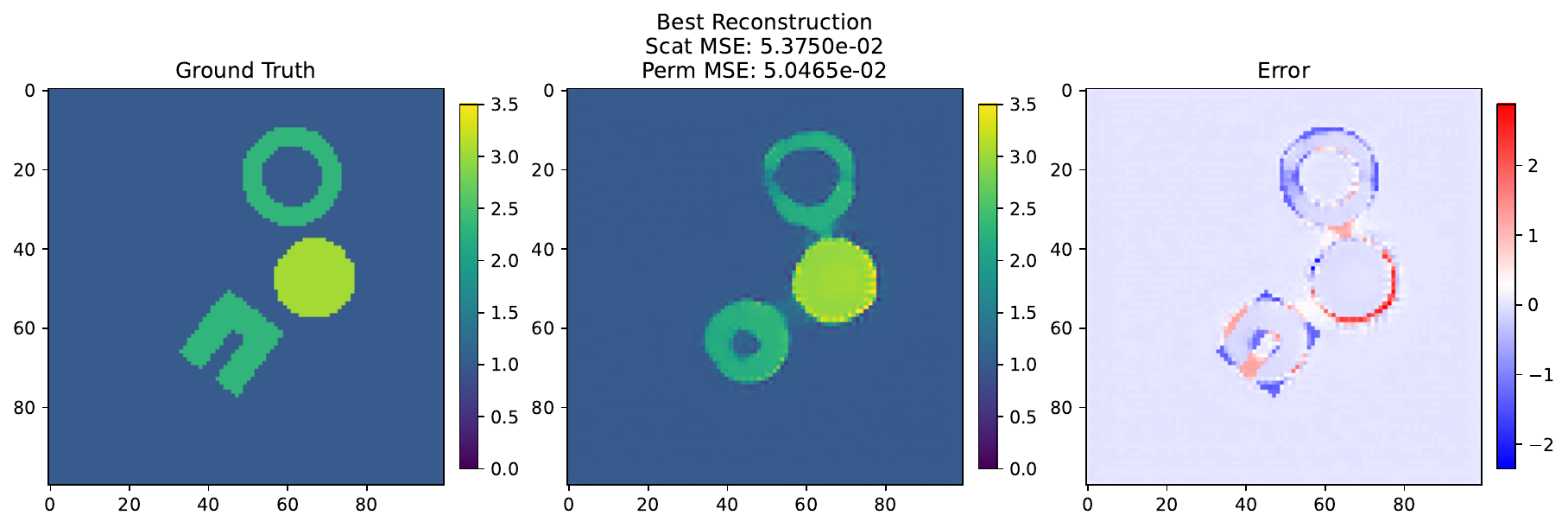}
  \includegraphics[width=0.9\columnwidth]{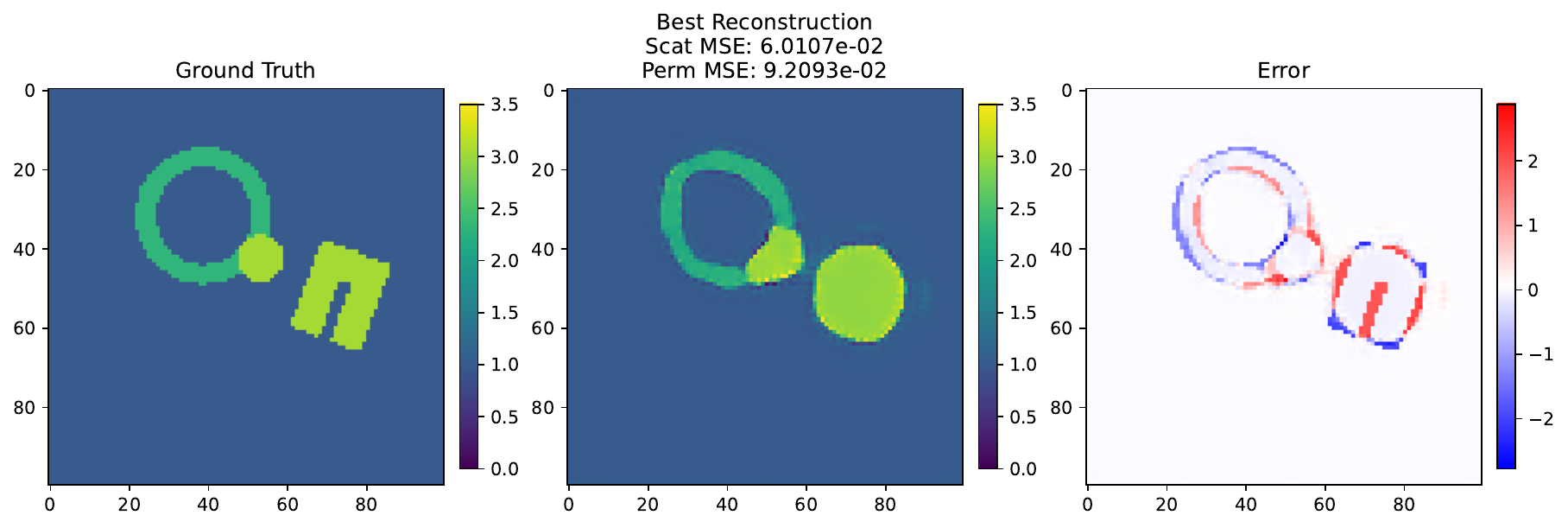}
  \includegraphics[width=0.9\columnwidth]{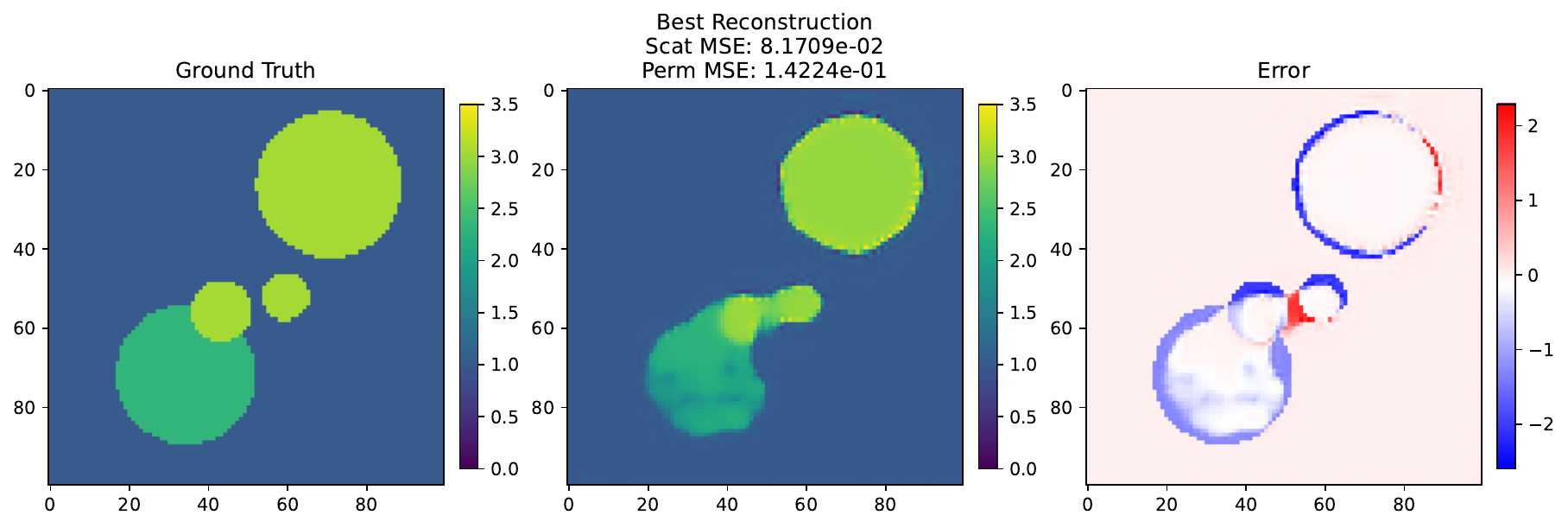}\\[1ex]
  \includegraphics[width=0.9\columnwidth]{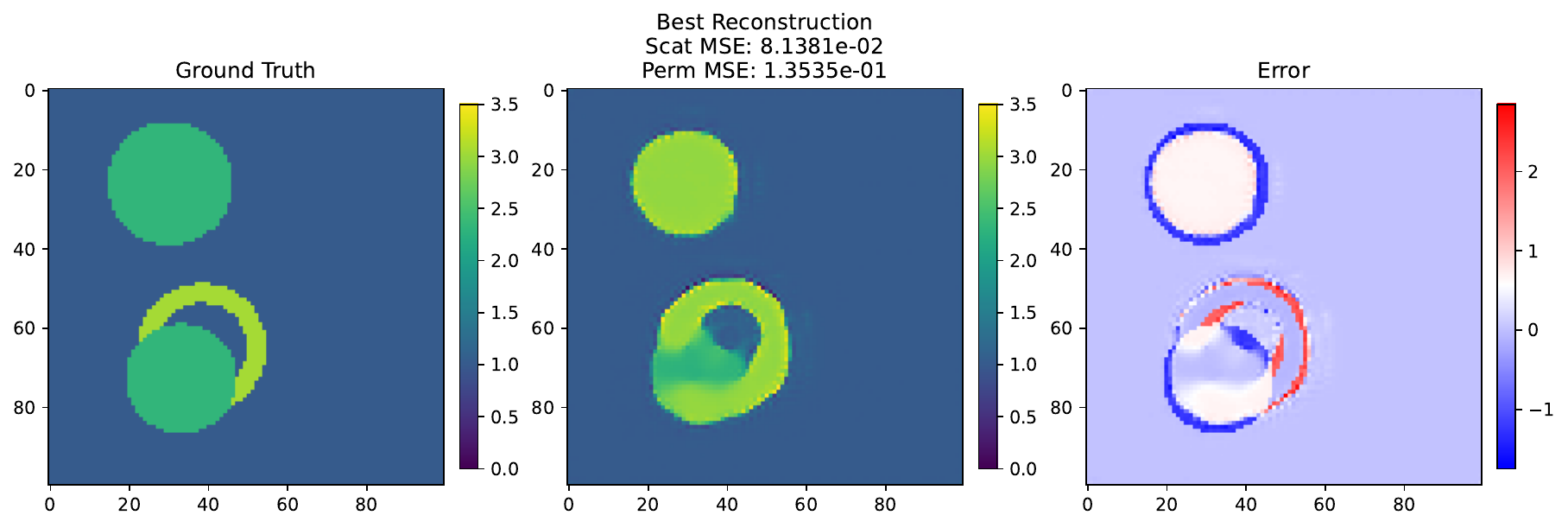}
  
  \caption{Evaluation of the proposed model using single frequency synthetic measurements for several representative samples from each category}
  \label{fig:dataset2_signle}
\end{figure}

\begin{figure}
  \centering
  \includegraphics[width=0.9\columnwidth]{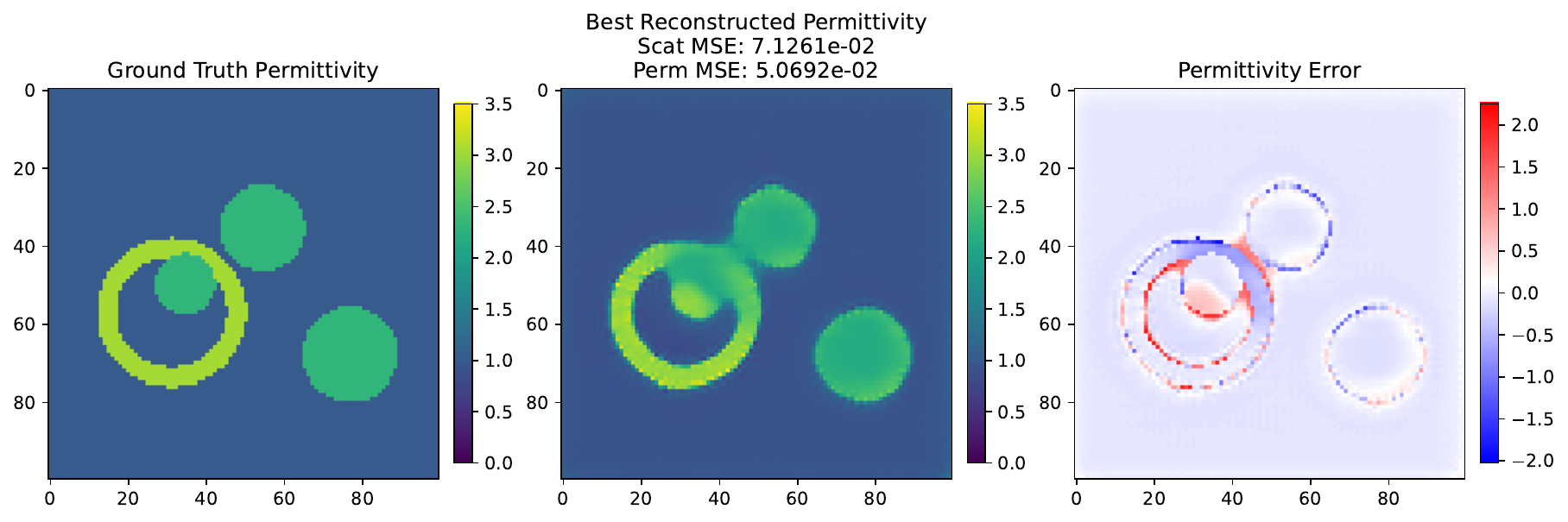}
  \hfill
  \includegraphics[width=0.9\columnwidth]{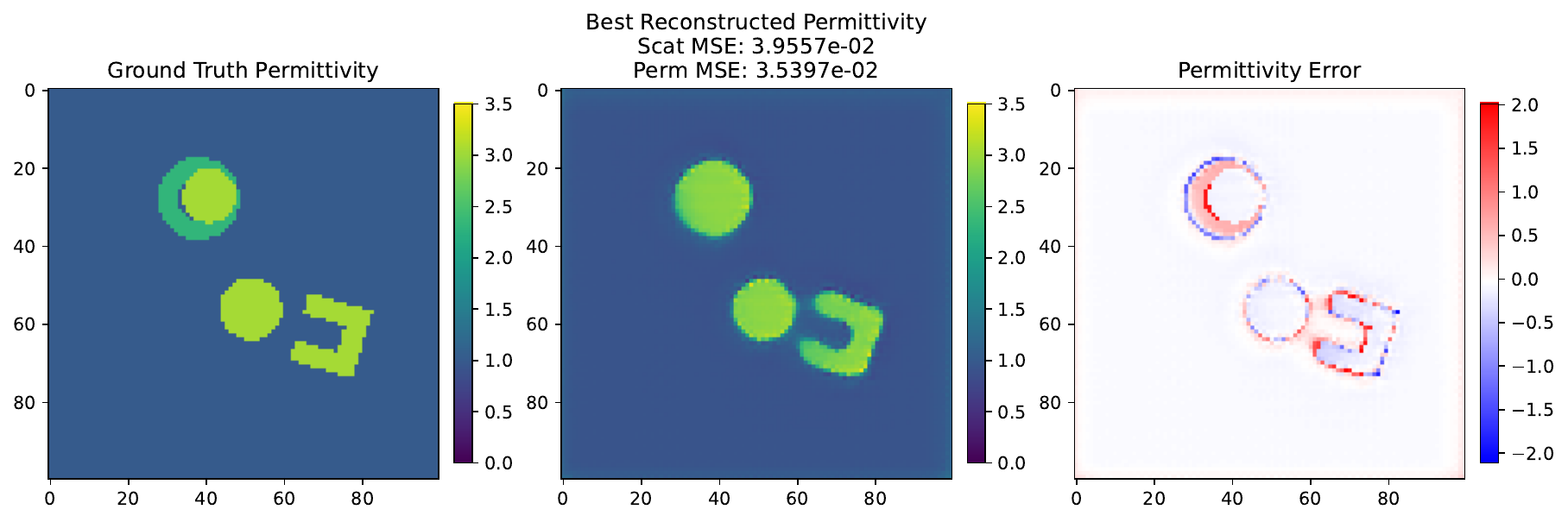}\\[1ex]
  \includegraphics[width=0.9\columnwidth]{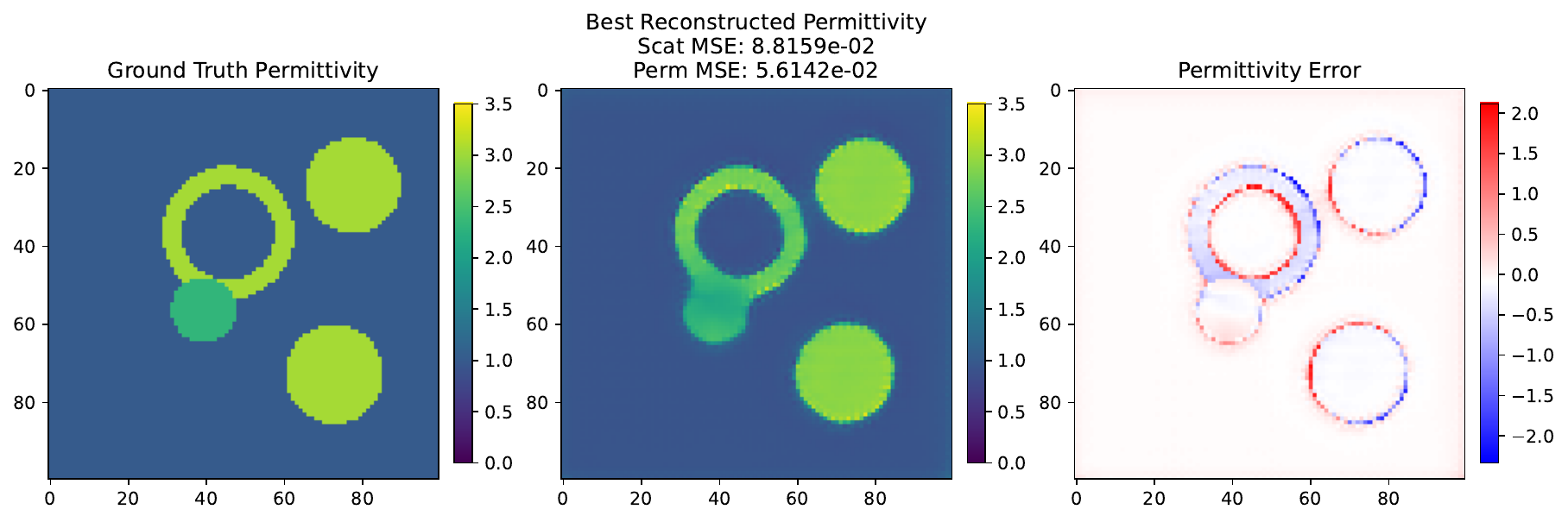}

    \includegraphics[width=0.9\columnwidth]{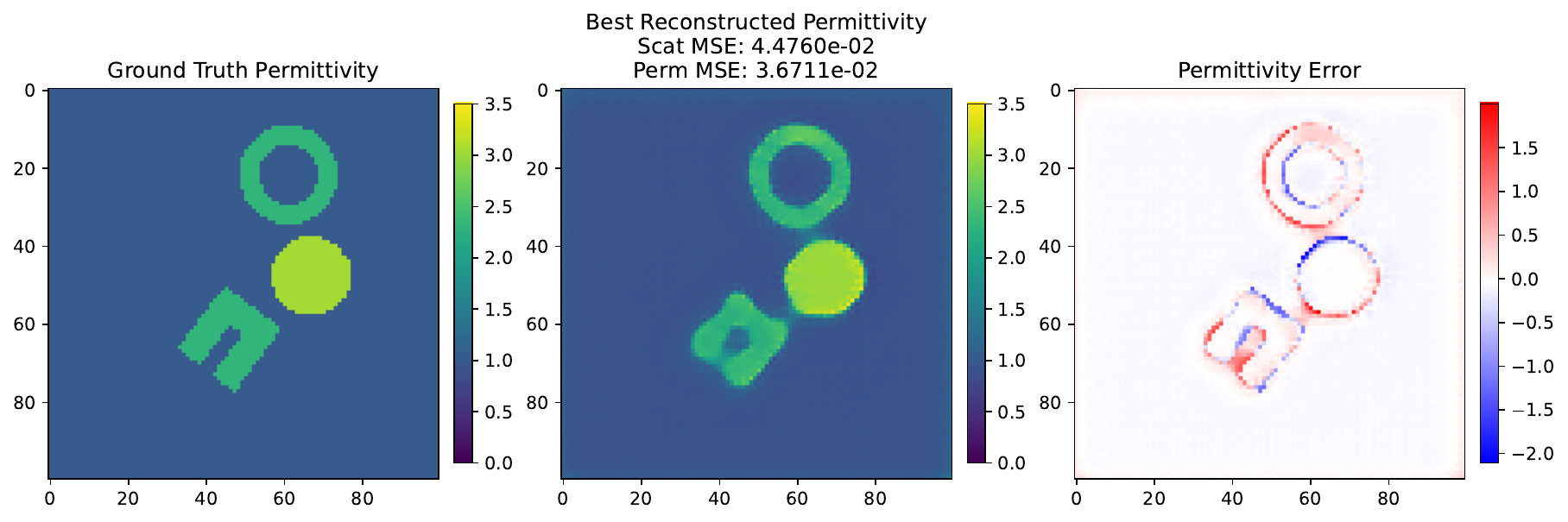}
  \includegraphics[width=0.9\columnwidth]{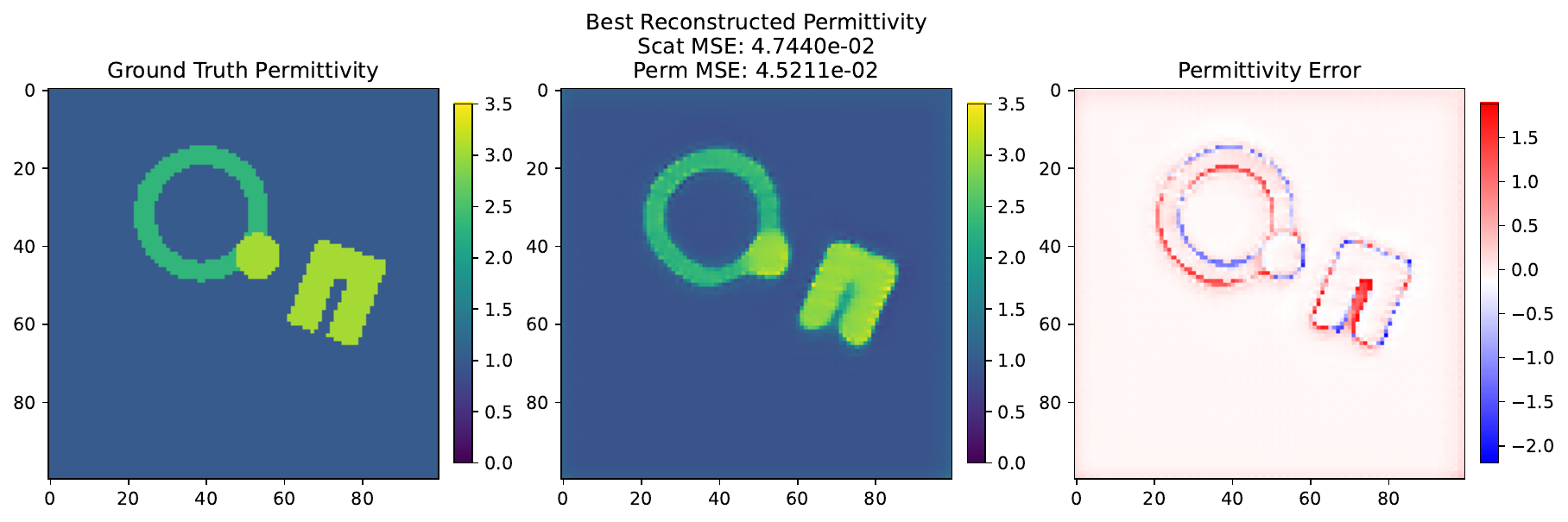}
  \includegraphics[width=0.9\columnwidth]{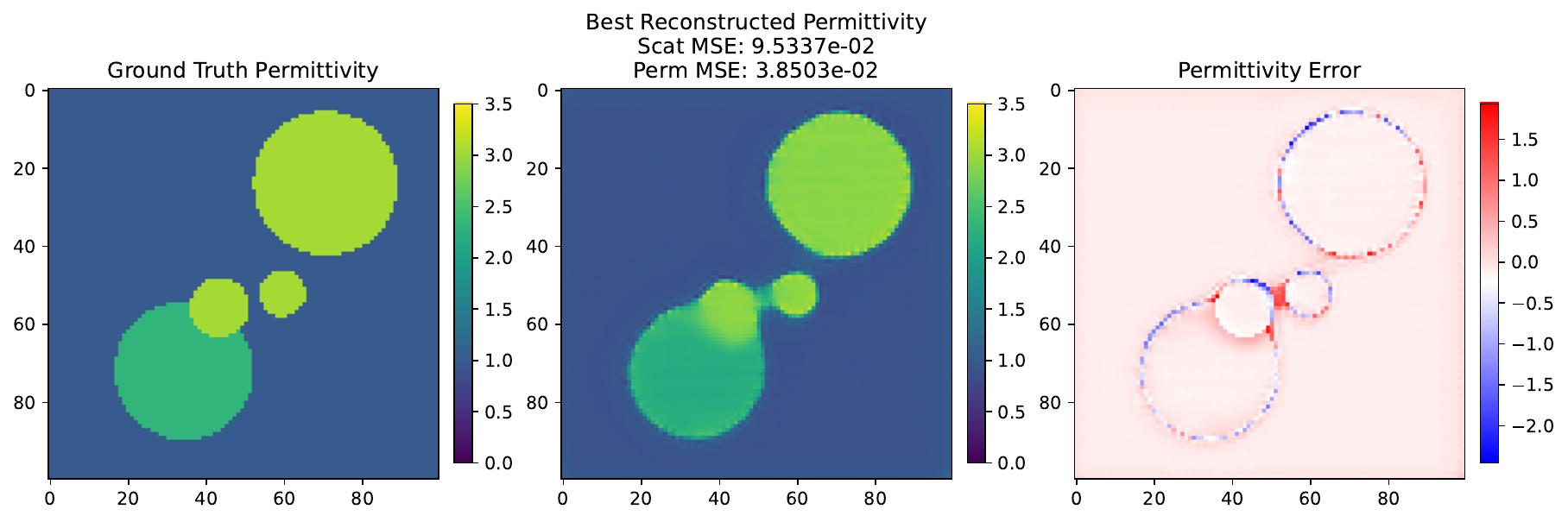}\\[1ex]
  \includegraphics[width=0.9\columnwidth]{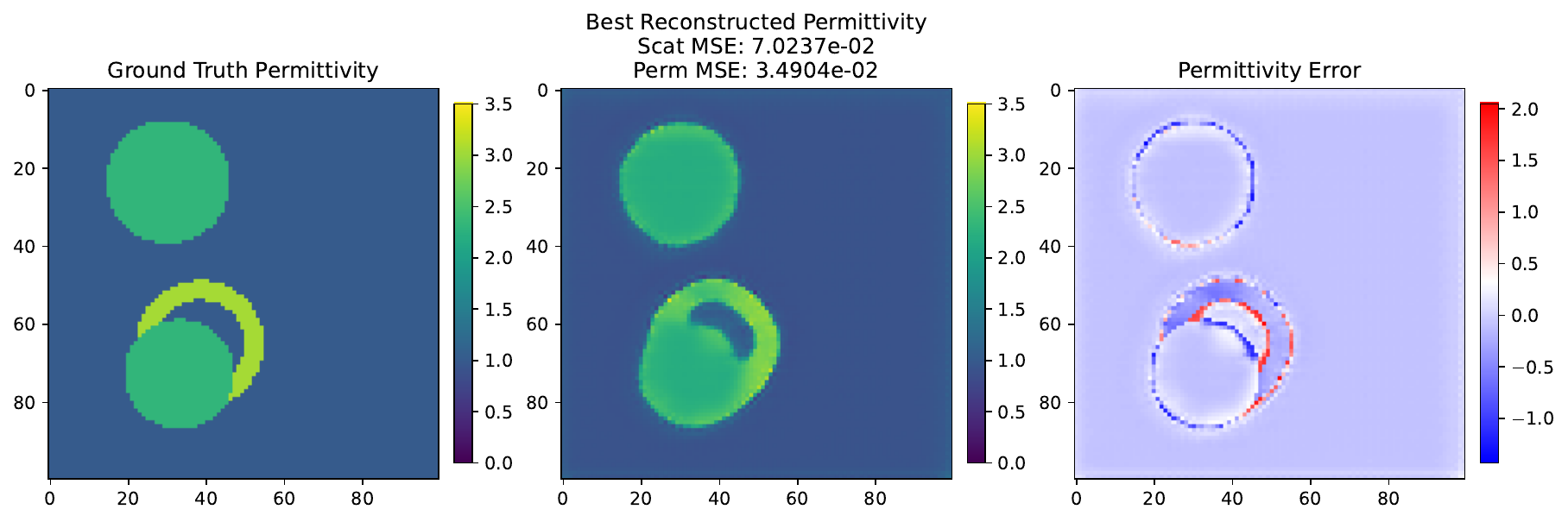}
  
  \caption{Evaluation of the proposed model using multi-frequency synthetic measurements for several representative samples from each category}
  \label{fig:dataset2_multi}
\end{figure}

Figure~\ref{fig:AE} qualitatively illustrates the compression error obtained when only the AE component of the model is applied to both datasets. While improving the AE architecture lies beyond the scope of this study, the results show that the compression-related error (1.39\%) accounts for approximately 25\% of the total reconstruction error (3.99\%). In an attempt to mitigate this effect, the AE performance was optimized through hyperparameter tuning, particularly by adjusting the relative weighting of loss terms to balance numerical accuracy with perceptual reconstruction quality. The reported results are for performance after this tuning was implemented.

\begin{table}[h!]
\centering
\caption{Comparison of AE compression performance averaged over 100 random test samples.}
\label{tab:ae_comparison}
\footnotesize
\begin{tabular}{|l|c|}
\hline
\textbf{Dataset} & \textbf{Average MSE} \\
\hline
DataSet1 & 0.71\% \\
\hline
DataSet2 & 1.39\% \\
\hline
\end{tabular}
\end{table}

\section{Conclusion}
This work presented a physics-informed conditional generative framework for solving the electromagnetic inverse scattering problem in microwave imaging. Leveraging the generative nature of the diffusion model, the proposed approach explicitly captures the inherent uncertainty and non-uniqueness of the inverse problem by producing multiple plausible reconstructions of the permittivity distribution. A forward-solver-based physical evaluation is then employed to select the most consistent reconstruction, ensuring that the final solution aligns with both the measurement data and the underlying electromagnetic physics.

The results demonstrated that the model can accurately reconstruct complex permittivity distributions from synthetic data, achieving stable and low-error reconstructions. When applied to experimental scattered-field measurements—despite being trained exclusively on synthetic data—the model successfully recovered key structural features, confirming its robustness and generalization capability across measurement domains. The proposed approach outperformed the state-of-the-art deep learning–based inverse solvers reported in \cite{Cathers2025}, achieving lower reconstruction error and improved structural fidelity in both synthetic and experimental evaluations. 
In order to ensure that our model was not overfitted to a limited number of object types—an issue commonly observed in previous studies—we developed a more diverse synthetic dataset. This expanded dataset enabled the model to better generalize to previously unseen geometries.  

Incorporating multi-frequency scattered-field data further improved reconstruction accuracy, yielding finer structural details and enhanced consistency across test samples. These findings are consistent with prior literature, underscoring the benefits of multi-frequency illumination in improving stability and resolution in microwave imaging.  

The results also revealed an increase in the reconstruction error attributable to the autoencoder (AE) component. Although optimizing the compression AE network lies beyond the scope of this study, preliminary hyperparameter tuning and loss-weight adjustments helped balance numerical accuracy and perceptual reconstruction quality.  

Overall, the proposed diffusion-based framework demonstrates strong potential for robust, data-driven microwave imaging. It achieves superior reconstruction quality relative to existing deep learning baselines and exhibits strong cross-domain generalization from synthetic to experimental data. Future work will extend this approach to full three-dimensional medical imaging and further refine the underlying AE architecture to enhance reconstruction fidelity and computational efficiency.

\section{Acknowledgment}
The authors acknowledge the help of Mr. Seth Cathers who supplied the forward solver.
\bibliographystyle{IEEEtran}
\bibliography{IEEEabrv,bibliography}

\vfill

\end{document}